\documentclass[12pt]{article}%
\usepackage{amssymb}
\usepackage{amsmath}
\usepackage{amsfonts}
\usepackage{graphicx}
\usepackage{color}
\usepackage{hyperref}
\providecommand{\U}[1]{\protect\rule{.1in}{.1in}}
\newtheorem{theorem}{Theorem}

\newtheorem{lemma}[theorem]{Lemma}

\renewcommand{\theequation}{\thesection.\arabic{equation}}

\newfont{\bbf}{cmbx12 scaled 1435}

\renewcommand{\theequation}{\thesection.\arabic{equation}}

\begin{document}

\title{{\LARGE Nonparametric identification of an interdependent value model with
buyer covariates from first-price auction bids}}
\author{Nathalie Gimenes\\Department of Economics\\PUC Rio\\Brazil
\and Emmanuel Guerre\\School of Economics\\University of Kent\\United Kingdom}
\date{October 2019}
\maketitle

\pagebreak

\begin{center}
\textbf{Abstract}
\end{center}

\bigskip

This paper introduces a version of the interdependent value model of Milgrom
and Weber (1982), where the signals are given by an index gathering signal shifters
observed by the econometrician and private ones specific to each bidders. The
model primitives are shown to be nonparametrically identified from first-price auction bids
under a testable mild rank condition. Identification holds for all possible signal values. This allows to consider a wide range of counterfactuals where this is important,  as
expected revenue in second-price auction. An estimation procedure
is briefly discussed.

\bigskip


\bigskip


\vfill

{\footnotesize The authors acknowledge useful discussions, comments and
encouragements from various conferences and seminars participants. Emmanuel Guerre would like to thank more specifically  Isabelle Perrigne and Quang Vuong for stimulating discussions. The comments of two Referees and the Editor Elie Tamer have been especially rich and useful to improve the paper. All
remaining errors are our responsibility. Both authors would like to thank the
School of Economics and Finance, Queen Mary University of London, for generous
funding.}

\bigskip

\pagebreak

\section{Introduction}

Most nonparametric identification results or empirical studies for first-price
auctions consider the private value case. See for instance the review papers
of Athey and Haile (2007) or Hendricks and Porter (2007). This is probably due
to the non identification result of Laffont and Vuong (1996), which states
that bids generated by interdependent values are also rationalized by private
values. This negative result has not prevented empirical work for the interdependent
case. In particular, Hendricks, Porter and Wilson (1994) consider an asymmetric common value model for drainage tracts, where an informed bidder bids for a neighbor tract and competes with non neighbor buyers.
Hendricks, Pinkse and Porter (2003) have used \textit{ex
post }values observed after the auction to test rational bidding in a common
value framework. They consider an application to wild cat leases, where bidders can commission 
seismic studies to evaluate the quantity of petrol or gas in the tracts.
Shneyerov (2006) has shown that the seller expected revenue
in a first or second price auction can still be identified. His results are applied to municipal bonds, which values are determined after the auction, when sold to some final investors. Paarsch (1992),
Haile, Hong and Shum (2003) and Compiani, Haile and Sant'Anna (2018) propose
to test whether bids are generated by a common or a private value model. Hong
and Shum (2002) restore identification using a parametric common value model.
Aradillas-L\'{o}pez, Gandhi and Quint (2013) use a set identification approach
in the more restrictive framework of correlated private values.

These papers fall in the interdependent value framework, where each bidder has
a possibly unknown value $V_{i}$ and observes a univariate private signal
$X_{i}$, $i=1,\ldots,n$. The parameters of interest are the joint signal
distribution and the \textit{valuation} or \textit{value functions}%
\begin{equation}
\Phi_{i}\left(  X_{1},\ldots,X_{n}\right)  =\mathbb{E}\left[  V_{i}\left\vert
X_{1},\ldots,X_{n}\right.  \right]  \label{Valfunc}%
\end{equation}
which are sufficient to compute most counterfactuals. 
In the private value case, the valuation only depends upon the bidder's signal. By contrast, in the interdependent case, the value of a given bidder can depend upon the signals of other bidders, creating so a network of interactions which is difficult to identify from first-price auction bids without further restrictions. 

In the symmetric case,
Laffont and Vuong (1996) show that the bids only identify
\[
\boldsymbol{V}_{i}\left(  x,x\right)  =\mathbb{E}\left[  V_{i}\left\vert
X_{i}=x,\max_{1\leq j\neq i\leq n}X_{j}=x\right.  \right]  .
\]
In particular, the ``pseudo'' private values $\boldsymbol{V}_{i}=\boldsymbol{V}_{i}\left(
X_{i},X_{i}\right)  $ generate equilibrium bids which are observationally
equivalent to the initial ones. As the private values $\boldsymbol{V}_{i}$ are independent and identically distributed (i.i.d) when the signals $X_i$'s are, it would be tempting to expect that revenue equivalence results valid under the symmetric independent private value paradigm extends to symmetric interdependent values. This is however misusing the observational equivalence results of Laffont and Vuong (1996) by ignoring bidder valuation dependence. In particular, it is known that ascending auctions, during which bidders can learn about their opponent's signals, generates a higher expected seller revenue than first-price auction. The optimal reserve price computed from the distribution of the pseudo private values
$\boldsymbol{V}_{i}$ is also unlikely to be identical with the one taking into account bidder's interdependence. As the valuation functions are typically positively correlated, so will be the bidder's participation decisions, suggesting the seller faces a higher risk of non participation under interdependent value than for independent private ones. Recovering the function $\Phi_i(\cdot)$ in (\ref{Valfunc}) and the signal distribution is therefore important from an auction design  perspective.

Functional restrictions have been used to restore identification within a symmetric framework. Hong and Shum (2002) have shown that a Gaussian Wilson model is identified.
Li, Perrigne and Vuong (2000)
consider an extension of the Wilson model where $V_{i}=V$ for all $i$, the signals satisfy
$X_{i}=V\varepsilon_{i}$. They assume that, for some unknown parameter $(\theta_0,\theta_1)$, $\boldsymbol{V}_{i}\left(  x,x\right)=\theta_0 x^{\theta_1}$  for identification and estimation purposes. F\'{e}vrier (2008) relaxes the latter assumption  and shows identification when the conditional support of  $X_i$ given the common value depends upon $V$.
He (2015) considers  $\Phi_i (x_1,\ldots,x_n) = \bar{x}$ for all $i$.

Alternatively, observable asymmetries generated by bidder specific variables can be used to obtain identification with such restrictions on $\Phi_i (\cdot)$ or $\boldsymbol{V}_i (\cdot)$. Somaini (2018) considers a valuation exclusion restriction, under which bidder $i$ valuation only depends upon a bidder specific shifter $Z_i$, which is observed by all bidders and the econometrician. This is sufficient to obtain identification for additive valuation function. The present paper explores another route, focusing on the signal, which is now supposed to be partly observable by all bidders and the econometrician. More specifically, all the signals $X_j$ appearing in (\ref{Valfunc}) can be decomposed into a $D$ dimensional common knowledge ``signal shifter'' $Z_j$ and a private  vector component $\gamma_{ij} (A_j)$, where $A_j$ is normalized to have a uniform distribution. These two components are combined using a linear index structure,  $X_j=Z_j^{\prime} \gamma_{ij} (A_j)$.\footnote{
			The index structure is general enough to include sieve approximation of a signal 
			function $x_{ij}\left(z_{j},A_{j}\right)$, provided $Z_j = [b_1(z_j),\ldots,b_D(z_j)]$ for a sieve 
			$\{b_d(\cdot)\}_{d=1}^{\infty}$ and $D$ growing with the sample size. Such extension is however out of the 
			scope of the present paper. 
	} In the latter  expression,   $\gamma_{ij}(\cdot)$ is a common knowledge slope function which may depend upon the identity of the considered bidder.
If so, $\gamma_{ij} (\cdot)$ incorporates a specific bidder $i$ fixed effect, or can be viewed as bidder $i$ belief about the slope $\gamma_{jj} (\cdot)$ of bidder $j$.

The common knowledge variable $Z$ can indicate better information or a higher
value.\footnote{Variation of the common knowledge slope functions $\gamma_{ij} (\cdot)$ across $i$ can also indicate a stronger bidder due to an unobserved variable which stay constant across the sample, or a fixed effect. For instance, each  bidder collection of $\gamma_{ij} (\cdot)$, $j=1,\ldots,n$ may take the two distinct values $\underline{\gamma}_{i} (\cdot) \leq \overline{\gamma}_{j} (\cdot)$, $j=1,\ldots,n$ indicating a weak or a strong bidder.}
For instance, Somaini (2018)  considers the bidder distance to the place where works of the auctioned contract has to be performed, which inverse can be both an indicator of increasing information or lower cost. Bidder capacity constraints can also affect the value of an auctioned contract.
Length of common tract borders can be used as a measure of information strength in the application of Hendricks et al. (1994.) 
The cost paid by bidders to acquire information on a gas or oil contents on an auctioned lease, as described in Hendricks, Pinske and Porter (2003), can also be used if common knowledge. 
Bidder reputation and experience, measured for instance by time spent in activities related with the auctioned good can again indicate a better information or a higher ability to process it for resale.

The variations of the continuous $Z_{j}$'s are used to
identify the function $\Phi_i(\cdot)$ in (\ref{Valfunc}) and the slope functions $\gamma_{ij}\left(
\cdot\right)  $, $j=1,\ldots,n$, from the bids observed in a first-price auction. The joint distribution of the private signals
is also identified assuming that a low signal does not prevent bidding and that bids increase with signal. Hence this interdependent value model is fully identified and can be
used for most usual counterfactual exercises. As our nonparametric
identification result uses local variation at given small value of the $Z_{j}%
$'s, the model is overidentified and can therefore be tested.

Athey and Haile (2002) have similarly considered good covariate variations for
ascending auctions with dependent private values. The nonparametric
identification result stated in Athey and Haile (2002, Theorem 5) relies on
order statistic properties that are not relevant here. The harder parameters
to identify are the slope functions $\gamma_{ij}\left(  \cdot\right)  $. In
the two bidder case, identification is obtained by differentiating with
respect to the signal $A_{i}$ and to the covariate $Z_j$. This gives a system of
differential equations with a unique solution, identifying so the slope
functions. Similar identification procedure, based on uniqueness of the
solution of a  differential equation, can be traced back to Elbers
and Ridder (1982) in the context of duration models, see also Abbring and van
der Berg (2003). The case where three bidders or more attend the auction is
more involved but proceeds similarly:  differentiating with respect to the
signal and covariate gives an integro-differential system, but it is shown that it also identifies the
slope functions.

The rest of the paper is organized as follows. The next section introduces our
interdependent value models and three illustrative examples. Section 3
presents our main nonparametric identification results. Section 4 discusses a
possible two stage estimation procedure. Section 5 concludes the paper and
proofs are grouped in Section 6.

\section{ Model, examples and assumptions\label{model}}

A single and indivisible object is sold to a known number $n\geq2$ of buyers
using a first-price auction. There is no reserve price and the seller accepts all nonnegative bids. The observations consist on the bidder
identities, bids $B_{j}$ and the signal shifters $Z_{j}$, $j=1,\ldots,n$ where $n$
is the total number of buyers. The next Section describes the model and
bidding strategies, and gives our key identification assumptions. The
framework focuses on a particular bidder, say bidder $i$, which valuation
function, or value, is a parameter of interest together with the joint signal
distribution introduced below.

\subsection{Valuation function and examples}

Buyers asymmetry is common knowledge and driven by individual $D$ dimensional
variables $Z_{j}$. Prior bidding, each buyers receive a private signal $A_{i}%
$. The joint distribution of the signals $A=\left(  A_{1},\ldots,A_{n}\right)
$ given $Z=(Z_1,\ldots,Z_n)$ is known to the buyers but not to the analyst. The marginal
distribution of each $A_{i}$ is normalized to be uniform over $\left[
0,1\right]  $. The  valuation function of  buyer $i$ is%
\begin{equation}
V_{i}\left(  A;Z\right)  =\Phi_{i}\left[  Z_{1}^{\prime}\gamma_{i1}\left(
A_{1}\right)  ,\ldots,Z_{n}^{\prime}\gamma_{in}\left(  A_{n}\right)  \right]
=\Phi_{i}\left[  Z,\Gamma_{i}\left(  A\right)  \right]  ,\label{MSMU}%
\end{equation}
$i=1,\ldots,n$, where $\Phi_{i}\left(  \cdot\right)  $ and $\Gamma_{i}\left(
A\right)  =\left(  \gamma_{i1}\left(  A_{1}\right)  ,\ldots,\gamma_{in}\left(
A_{n}\right)  \right)  $ are unknown parameters of interest.\footnote{This specification can easily be extended to allow for an auction specific variable $Z_0$, ie
$
V_{i}\left(  A;Z,Z_0\right)  
=
\Phi_{i}
\left[
\left.  
Z_{1}^{\prime}\gamma_{i1}\left(
A_{1}|Z_0\right)  ,\ldots,Z_{n}^{\prime}\gamma_{in}\left(  A_{n}|Z_0\right)
\right| Z_0
\right]$, by conditioning on $Z_0$. Identification of this model follows from applications, for each value of $Z_0$, of Theorems \ref{MSMidentn2} and \ref{MSMidentn3} below.}
In this specification, the index
$Z_{j}^{\prime}\gamma_{ij}\left(  A_{j}\right)  $ can be viewed as a
\emph{mixed signal} combining observable and unobservable components, and
replaces the signal $X_{i}$ from (\ref{Valfunc}). The function $\Phi_{i}$
combines these signals and reveals the interactions of the other bidders with
$i$, in terms of value. Examples are as follows, for which $\Gamma_{i}\left(
\cdot\right)  =\Gamma\left(  \cdot\right)  $ is common to buyers.

\begin{itemize}
\item \textbf{Additive valuation model}. Bidder $j$ observes a component
$Z_{j}^{\prime}\gamma_{j}\left(  A_{i}\right)  $ of the total valuation of the
auctioned good, which is weighted with a weight $\pi_{ij}$ by bidder $i$ in
her value function%
\[
V_{i}\left(  A;Z\right)  =\sum_{j=1}^{n}\pi_{ij}Z_{j}^{\prime}\gamma
_{j}\left(  A_{j}\right)  \text{ with }\pi_{ij}\geq0
\]
in which case $\Phi_{i}\left(  x_{1},\ldots,x_{n}\right)  =\sum_{j=1}^{n}%
\pi_{ij}x_{i}$. He (2015) obtains identification in the symmetric case
$V_{i}\left(  A;Z\right)  =\sum_{j=1}^{n}\gamma\left(  A_{j}\right)  /n$, but
asymmetric specifications can be more relevant for applications. For instance
the auctioned good can be a piece of land expected to contain some resources,
as in mineral rights auctions. In this type of auction, bidders exploit
similar lots which can be adjacent or have similar characteristics to the
auctioned one, a observable information that can be recorded in $Z$. The
signal $A_{i}$ is private to bidder $i$, being for instance the outcomes of
her lot. A $\pi_{ij}$ set to $0$ may indicate that bidders $i$ and $j$ would
use the lot for different purposes, so that information of $i$ is not relevant
for $j$ and vice versa. 

\item \textbf{A simplified auction with resale. }Suppose each bidder is tied
with a final buyer to whom he can sell for sure the good at a price $\pi
_{ii}Z_{i}^{\prime}\gamma_{i}\left(  A_{i}\right)  $ if she wins the auction.
After the auction, the winner can sell the good to the other final buyers at a
price $\pi_{ij}Z_{j}^{\prime}\gamma_{j}\left(  A_{j}\right)  $. This gives the
value functions
\[
V_{i}\left(  A;Z\right)  =\max_{i=1,\ldots,n}\left\{  \pi_{ij}Z_{j}^{\prime
}\gamma_{j}\left(  A_{j}\right)  \right\}
\]
in which case $\Phi_{i}\left(  x_{1},\ldots,x_{n}\right)  =\max_{i=1,\ldots
,n}\left\{  \pi_{ij}x_{i}\right\}  $.

\item
\textbf{A non-Gaussian and asymmetric Wilson model}. Suppose the value $V$ of the good
	is not observed by the buyers, who receive instead a noisy signal which
	accuracy is bidder specific and common knowledge. A possible parameter of interest is the quantile
	function $\gamma_{0}\left(  \cdot\right)  $ of $V$. Hence for a uniform
	$A_{0}$, $V=\gamma_{0}\left(  A_{0}\right)$, which can also be written as
	\[
	V=\gamma_{0}\left(  F (\nu) \right) 
	\]
	where $\nu$ is a standard normal and $F(\cdot)$ its cdf. In the standard Wilson model, the value is normal, ie $V=\nu$ up to a scale parameter, and assuming that the distribution of $V$ is unknown introduces additional complications. The signal structure resembles the standard Wilson model, the signal of bidder $i$ with observed accuracy $\sigma_i$ for $\nu$ being
	\[ 
	   \nu_i = \nu + \sigma_i \varepsilon_i
	\]
	with independent $\nu$, $\varepsilon_1,\ldots,\varepsilon_n$ drawn from the standard normal distribution. Note that the information carried by the signal $\nu_i$ is equivalent to the uniform signal
	\[
	    A_i = F \left( \frac{\nu_i}{\sqrt{1+\sigma_i^2}}\right)
	\]
	It follows that the distribution of $\nu$ given $A_1,\ldots,A_n$ is
	\[
	\mathcal{N}
	\left(
	\frac{\sum_{i=1}^{n} \frac{\nu_i}{\sigma_i^2}}{1+\sum_{i=1}^{n} \frac{1}{\sigma_i^2}}
	,
	\frac{1}{1+\sum_{i=1}^{n} \frac{1}{\sigma_i^2}}
	\right)
	=
	\mathcal{N}
	\left(
	\frac{\sum_{i=1}^{n} \frac{\sqrt{1+\sigma_i^2}}{\sigma_i^2}F^{-1} (A_i)}{1+\sum_{i=1}^{n} \frac{1}{\sigma_i^2}}
	,
	\frac{1}{1+\sum_{i=1}^{n} \frac{1}{\sigma_i^2}}
	\right)
	\] 
	Set $Z_i=\frac{\sqrt{1+\sigma_i^2}}{\sigma_i^2}$, which gives $\frac{1}{\sigma_i^2} = \left( Z^2_i + \frac{1}{4} \right)^{1/2}-\frac{1}{2}$, so that a large $Z_i$ means a small signal variance $\sigma_i^2$.  The value function
	$V(A;Z)
	=
	\mathbb{E}
	\left[
	V
	\left| A,Z \right.
	\right]$
	is identical across buyers and
	satisfies, for $\Sigma^2 (Z) = 1+\sum_{i=1}^{n} \frac{1}{\sigma_i^2}$,
	\begin{align*}
	V(A;Z)
	&
	=
	\frac{1}{\sqrt{2 \pi}}
	\int
	\gamma_0
	\left(
	\frac{\sum_{i=1}^{n} Z_i F^{-1} (A_i)}{\Sigma^2 (Z)}
	+
	\Sigma (Z) t
	\right)
	\exp \left(-\frac{t^2}{2} \right)
	dt
	\\
	& =
	\Psi 
	\left(  
	  \sum_{i=1}^{n}Z_{i}F^{-1}\left(  A_{i}\right), \Sigma (Z) 
	\right)
	\end{align*}
	which, conditionally on $\Sigma (Z)$, falls in the considered framework. Note that, for $Z_j=\infty$, the variance $\sigma_j$ vanishes so that bidder $j$ knows the value $V$. Little algebra then gives that 
	\begin{equation}
	\lim_{Z_{i}\rightarrow+\infty} V\left(  A;Z\right)  =\gamma_{0}\left(
	A_{i}\right)  ,\text{ for any } i=1,\ldots,n,\label{Wilson}%
	\end{equation}
	showing that the quantile function $\gamma_0 (\cdot)$ is identified from the value function under a large support assumption for the bidder covariate.
\end{itemize}
\bigskip

As detailed below, the revisited Wilson model is easier to identify than the
other Examples, due to the possibility of perfect information. We now detail our main Assumptions for the value specification. In the sequel, we shall focus on the identification of buyer $i$ valuation function. Let $\mathcal{I}$ be the set of active signals, that is the signal affecting the valuation of bidder $i$, or in other words, the smallest subset of $\left\{  1,\ldots ,n\right\}  $ such that
\begin{equation}
	\Phi_{i}\left(  x_{1},\ldots,x_{n}\right)  =\Phi_{i}\left[  x_{j}%
	,j\in\mathcal{I}\right]  . \label{Phisupp}%
\end{equation}

\bigskip

\textbf{Assumption Z. }\emph{The variable }$Z=\left(  Z_{1},\ldots,Z_{n}\right)  $ \emph{has support }$\mathbb{R}_{+*}^{D} \times\cdots\times\mathbb{R}_{+*}^{D}=\mathcal{Z}$.

\bigskip

\textbf{Assumption A. }\emph{The buyer signal }$A_{j}$ \emph{is uniform over
}$\left[  0,1\right]  $ \emph{given }$Z$, $j=1,\ldots,n$\emph{. The joint
p.d.f }$c\left(  \cdot|Z\right)  $ \emph{of }$A=(A_1,\ldots,A_n)$ \emph{given }$Z$ \emph{is
strictly positive and continuously differentiable with respect to the signal
and the covariate}.

\bigskip

\textbf{Assumption G. }\emph{ The slope }$\gamma_{ij}\left(  \cdot\right)  $
	\emph{are continuously differentiable over }$\left[  0,1\right]  $ \emph{with
		nonnegative and nondecreasing entries, at least one being strictly increasing.  \textit{In addition, one of the following terminal and initial conditions holds}:\textit{(i) $\gamma_{ij} (1) \neq 0$ for all $j$ in $\mathcal{I}$ or (ii) $\gamma_{ij} (0) \neq 0$ for all $j$ in $\mathcal{I}$}}

\bigskip

\textbf{Assumption P. }
\emph{The set $\mathcal{I}$ is not empty. The function }$\Phi_{i}\left(  x_{1},\ldots
	,x_{n}\right)  $ \emph{maps }$\mathbb{R}_{+}^{n}$ \emph{into }$\mathbb{R}_{+*}$ \emph{and is twice continuously differentiable with partial derivatives
}$\frac{\partial\Phi_{i}\left(  \cdot\right)  }{\partial x_{j}}%
>0$ \emph{over }$\mathbb{R}_{+}^{n}$ \emph{for all $j$ in $\mathcal{I}$.}

\bigskip

Assumption Z imposes signal shifter linear
independence and rules out discrete  entries to allow differentiation.
Constant entries are also ruled out as it can cause identification loss as in Laffont and Vuong (1996).
That the entries of $Z$ can be unbounded is a simplifying assumption that can be weakened, but allows here to identifies $\Phi_{1}(\cdot)$ over its unbounded definition domain. That the vector $Z$ can go to $0$ is used to identify the initial or terminal values of the $\gamma_{ij} (\cdot)$'s later on.

Assumption A includes a standard normalization of the
conditional signal distribution, which are assumed to be uniform.  This
permits the use of the quantile approach of Gimenes and Guerre (2019), see 
Lemma \ref{Bidder2} below. Note however that the signal vector $(A_1,\ldots,A_n)$ can depend upon $Z$.

Assumption G imposes smooth and bounded mixed signals
$Z_{j}^{\prime}\gamma_{ij}\left(  \cdot\right)$.\footnote{Assumption G rules out the
revisited Wilson model example, for which $\gamma_{ji}\left(  \cdot\right)  $
is infinite at $0$ or $1$. But Assumption G is only needed to identify the
unknown $\gamma_{ji}\left(  \cdot\right)  $, and therefore not for this
example since its $\gamma_{ij}\left(  \cdot\right)  $ are known.} 
The terminal slope condition G-(i) will be used when $n=2$, while the initial one G-(ii), which implies G-(i) when the slope entries are strictly increasing, is used for $n=3$.
Assumption P
requests a valuation function which is strictly increasing with respect to its
argument and smooth. Assumptions G and P together imply that the value
function $V_{i}\left(  A;Z\right)  $ increases with each private signal
$A_{j}$, as assumed in Milgrom and Weber (1982). In the sequel, we shall focus
on the identification of buyer $i$ valuation function, or equivalently of the
pair $\left[  \Phi_{i}\left(  \cdot\right)  ,\Gamma_{i}\left(  \cdot\right)
\right]  $ up to a scaling normalization, as  $\left[  \Phi_{i}\left(  \cdot/\lambda\right)  ,\lambda\Gamma_{i}\left(  \cdot\right)
\right]  $ gives the same valuation function for all $\lambda>0$. A convenient normalization used in the proofs of the main results is
\begin{equation}
\frac{\partial \Phi_i \left(0,\ldots,0\right)}{\partial x_j}
=1
\text{ for all $j$ in $\mathcal{I}$.}
\label{Normphi}
\end{equation}

\subsection{Bidding strategies}

Our identification results are based on high level bidding assumptions
inspired by the Bayesian Nash Equilibrium framework. It is assumed that bids
depend upon $Z$ and bidder private signals
\[
B_{j}=s_{j}\left(  A_{j};Z\right) > 0 ,\quad j=1,\ldots,n,\quad Z\in\mathcal{Z}.
\]
In what follows, $\varphi^{(1)}(\cdot;z)$ is the partial derivative $\varphi^{(1)}(\alpha;z)=\frac{\partial}{\partial \alpha} \varphi(\alpha;z)$ with respect to the quantile level.

\bigskip

\textbf{Assumption S. }\textit{For each }$Z$ \textit{in }$\mathcal{Z}$
\textit{(i)} \textit{the bidding strategy }$s_{i}\left(  \cdot;Z\right)  $
\textit{satisfies the best response condition}%
\begin{equation}
	s_{i}\left(  \alpha;Z\right)  \in\arg\max_{b>0}\mathbb{E}\left[  \left(
	V_{i}\left(  A;Z\right)  -b\right)  \mathbb{I}\left\{  b\geq\max_{1\leq j\neq
		i\leq n}B_{j}\right\}  \left\vert A_{i}=\alpha,Z\right.  \right].  \label{Brc}%
\end{equation}	
\textit{(ii) The initial condition  $s_{1}\left(  0;Z\right)  =\cdots=s_{n}\left(
		0;Z\right)  $ holds.}
\textit{(iii) The terminal condition  $s_{1}\left(  1;Z\right)  =\cdots=s_{n}\left(
		1;Z\right)  $ holds.} 
\textit{(iv) For each} $j=1,\ldots,n$, $s_{j}\left(  \cdot;Z\right)  $
\textit{is twice continuously differentiable with }$s^{(1)}_{j}\left(  \cdot;Z\right)  >0$ \textit{over }$\left[
	0,1\right]  $.

\bigskip

As in the preceding section, the focus is on identification of the valuation function of bidder $i$, so that the best
response condition (\ref{Brc}) in Assumption S-(i) only concerns this specific
bidder: the other bidders do not need to use a best response bidding strategy. Note however that, apart the initial condition of S-(ii), most of Assumption S is inspired by the Bayesian Nash Equilibrium framework where all the bidders use a best response bidding strategy.

Assumption S-(iv) is used in particular for the joint signal distribution, see Lemma \ref{Bidder2}, together with S-(ii,iii) to get its identification over the whole support $[0,1]^n$. The
monotonicity condition in S-(iv) is standard in the Econometrics of auctions
when assuming bids from the Bayesian Nash Equilibrium. For affiliated signal
and valuation functions $V_{i}\left(  A;Z\right)  $ increasing with respect to
each signals, Reny and Zamir (2004) have established existence, but not
uniqueness, of a Bayesian Nash equilibrium with increasing bidding strategies.
In the two bidder case, Lizzeri and Persico (2000, Appendix) have studied
uniqueness, strict monotonicity and smoothness of the optimal bidding
strategies as in S-(iv). But these assumptions may not hold for bids not generated by a Bayesian Nash equilibrium. Discontinuities in the strategy $s_j (\cdot|Z)$ should generate flat parts in the cumulative distribution function of $B_j$ or discontinuities for its pdf. A non monotonic but differentiable $s_j (\cdot|Z)$ can lead to a diverging pdf, so that both are in principle testable.

The common terminal condition in Assumption S-(iii) is from Lizzeri and Persico (2000), who established it in the two bidder case under the Bayesian Nash Equilibrium framework. Intuitively, if a group of bidders have a common terminal bid larger than their opponents, they can increase their profit by slightly decreasing their terminal bids and still be sure to win the auction. Hence all the bidders should have the same $s_j(1;Z)$ as assumed in S-(iii). This condition can be tested in principle, because if S-(iii) does not hold for a given $Z$, 
there is a dominated bidder, say $i_0$, with bid support upper bound $\overline{b}_0 (Z)$ 
such that  
$
\mathbb{P}
\left(
\left.
B_{i_0}
\leq 
\overline{b}_0 (Z)
\right|
Z
\right)
=
1
$
and
$
\mathbb{P}
\left(
\left.
\max_{1 \leq i \leq n}
B_i
>
\overline{b}_0 (Z)
\right|
Z
\right)
>
0
$
.

The initial condition in Assumption S-(ii) does not necessarily hold in the Bayesian Nash Equilibrium framework, especially with asymmetric bidders. It ensures that there is no bidder who would lose the auction
with probability 1 given a too low signal value. As a consequence, the best-response characterization (\ref{Brc}) of the bidding strategy implies the first order condition (\ref{B2U}) in Lemma \ref{Ident2} below, which is key for identification. Another important econometric role of the initial condition S-(ii) together with the terminal one S-(iii) is to allow for identification of the slope functions $\gamma_{ij} (\cdot)$ over the whole quantile interval $[0,1]$, therefore avoiding censoring. 
As Assumption S-(iii), Assumption S-(ii)  is in principle testable.

\subparagraph{Strategy common initial condition and Bayesian Nash equilibrium.}
For bids from Bayesian Nash equilibrium, additional restrictions may be needed to ensure that the common initial condition S-(ii) holds, as discussed now. Assumption S-(ii) typically fails when the initial valuations $V_i (0;Z)$ differ across bidders. If the $V_i (0;Z)$ are identical, say equal to $V(0;Z)$,  it must hold for all $i$
$s_i(0;Z) = V(0;Z)$ under Assumption S-(iv).\footnote{To see this, let 
$
\underline{\alpha}_i (Z)= 
\inf 
\left\{
	\alpha \in [0,1]; 
	\mathbb{P}\left(  s_i(A_i;Z) \text{ wins and }A_i \geq \alpha | Z \right) > 0
\right\}
$ be the signal threshold above which bidder $i$ has a non trivial probability to win the auction. Under Assumption S-(iv), it must hold
that
$s_1 (\underline{\alpha}_1 (Z);Z) = \cdots = s_n (\underline{\alpha}_n (Z);Z)$. Arguing as for (\ref{U1I2}), (\ref{U1I3}) and (\ref{B2U}) give
\[
s_j (\underline{\alpha}_j (Z);Z)
=
\Phi_j
\left[
Z_1^{\prime}
\gamma_{j1} (\underline{\alpha}_1 (Z))
,
\ldots,
Z_n^{\prime}
\gamma_{jn} (\underline{\alpha}_n (Z))
\right]
\]
which is strictly larger than the common $V(0;Z)$ if one of the $\underline{\alpha}_i (Z)$ is strictly larger than $0$ by Assumptions G and P. Suppose now that one of $\underline{\alpha}_i (Z)$ is strictly larger than $0$, say $\underline{\alpha}_1 (Z)>0$ assuming it is the unique one for the sake of brevity. It follows that bidder $i=2$ expected profit given $A_2=\alpha$ is equivalent to, when $\alpha$ goes to $0$
\[
C
\alpha^{n-1}
\int_0^{\underline{\alpha}_1 } 
\left\{
\Phi_2
\left[
Z_1^{\prime}
\gamma_{21} ( \alpha)
,
Z_2^{\prime}
\gamma_{22} (0)
,
\ldots,
Z_n^{\prime}
\gamma_{2n} (0)
\right] 
-
\Phi_2
\left[
Z_1^{\prime}
\gamma_{21} (\underline{\alpha}_1 (Z))
Z_2^{\prime}
\gamma_{22} (0)
,
\ldots,
Z_n^{\prime}
\gamma_{2n} (0)
\right]
\right\}
d \alpha
<0
\]
which is not possible as bidder 2 would achieve a non negative expected profit making a bid close to $V(0;Z)$ when $A_2$ goes to $0$. Hence it must hold that
$\underline{\alpha}_1 (Z)=0$.
}
Hence Assumption S-(ii) holds if
\begin{equation}
V(0;Z)
=
\Phi_1 \left(Z_1^{\prime} \gamma_{11} (0), \ldots, Z_n^{\prime} \gamma_{1n} (0)\right)
=
\cdots
=
\Phi_n \left(Z_1^{\prime} \gamma_{n1} (0), \ldots, Z_n^{\prime} \gamma_{nn} (0)\right).
\label{Vi0Sii}
\end{equation}
Assuming, for the sake of simplicity, that the set $\mathcal{I}$ of all bidders is $\{1,\ldots,n\}$  and the normalization (\ref{Normphi}), (\ref{Vi0Sii}) holds when  $\gamma_{j1} (0) = \cdots = \gamma_{jn} (0)$ for all $j$ and, if these slope do not vanish,  $\Phi_1 (\cdot) = \cdots = \Phi_n (\cdot)$. So a sufficient condition for Assumption S-(ii) is the following:

\bigskip

\noindent	
\textbf{Condition BNE.} \emph{One of the two restrictions holds:}
\begin{enumerate}
	\item 
	\emph{Common initial value:
	$\Phi_1 (0,\ldots,0)=\cdots=\Phi_n (0,\ldots,0)$ and $\gamma_{ij} (0)=0$ for all $i,j=1,\ldots,n$;}
	\item 
	\emph{Common $\Phi_i (\cdot)$ and initial slope:
	$\Phi_i (x_1,\ldots,x_n)=\Phi (x_1,\ldots,x_n)$ and $\left(\gamma_{i1} (0),\ldots,\gamma_{in} (0) \right)= \left(\gamma_{1} (0),\ldots,\gamma_{n} (0) \right)$ for all $i$.}
\end{enumerate}

\bigskip

Condition BNE-(i) allows for more asymmetry, with a set $\mathcal{I}$ that can vary across bidders. The common initial condition holds with $\Phi_i (0)=0$ in the three examples considered earlier. By contrast, the common $\Phi_i (\cdot)$ condition of BNE-(ii) is more demanding: it holds in the Wilson model examples but requests, in the two first examples, that the weights do not depend upon bidder identity, $\pi_{ij}=\pi_j$. More generally, Condition BNE-(ii) forces the set $\mathcal{I}$ of active bidders to be identical across bidders. Because the slope $\gamma_{ij} (\cdot)$ can be very small, this may  nevertheless be flexible enough to mimic any set $\mathcal{I}$ of active signals.

\subsection{Rank condition}

Our main identification condition is a rank condition stated in Assumption
I below. Let
\[
G_{j}\left(  b|Z\right)  =\mathbb{P}\left(  B_{j}\leq b|Z\right)
\]
be the c.d.f of $B_{j}$ given $Z$, $B_{j}\left(  \alpha|Z\right)  =G_{j}%
^{-1}\left(  \alpha|Z\right)  $ be the associated conditional quantile
function and set%
\[
G_{j}B_{i}\left(  \alpha|Z\right)  =G_{j}\left[  B_{i}\left(  \alpha|Z\right)
|Z\right]  .
\]
An important consequence of Assumption S-(ii,iii) and Lemma \ref{Bidder2}-(ii)
below, which states that $s_i (\cdot|Z)=B_i(\cdot|Z)$, is that $G_{j}B_{i}\left(  0|Z\right)  =0$ and $G_{j}B_{i}\left(
1|Z\right)  =1$ for all $j$. Recall that $\mathcal{I}$ is the smallest subset satisfying
$
\Phi_{i}\left(  x_{1},\ldots,x_{n}\right)  =\Phi_{i}\left[  x_{j}%
,j\in\mathcal{I}\right]  $.
Recall $Z_{k}^{\prime}=\left(  Z_{1k},\ldots,Z_{Dk}\right)  ^{\prime}$ and
let 
$
\partial_{Z_{k}} G_{j}B_{i}\left(  \alpha|Z\right)
$
be the gradient column vector
\[
\partial_{Z_{k}} G_{j}B_{i}\left(  \alpha|Z\right) =\left[
\frac{\partial G_{j}B_{i}\left(  \alpha|Z\right)  }{\partial Z_{1k}}%
,\ldots,\frac{\partial G_{j}B_{i}\left(  \alpha|Z\right)  }{\partial Z_{Dk}%
}\right]  ^{\prime}.
\]

\bigskip

\textbf{Assumption I. }
\emph{Let $\mathcal{I}$ be as in (\ref{Phisupp}) and suppose $\mathcal{I}\neq \{i\}$. For all $j$ in $\mathcal{I}\setminus \{i\}$, $G_{j}B_{i}\left(  \alpha|Z\right)$ 
is twice continuously differentiable with respect to $\left(  \alpha,Z\right)$ in $\left[  0,1\right]
\times\mathcal{Z}$. For
		all $\left(  \alpha,Z\right)  $ of $\left[  0,1\right)
	\times\mathcal{Z}$, the $\left(\mathrm{Card} (\mathcal{I})-1\right) \times \left(\mathrm{Card} (\mathcal{I})-1\right)$ matrix with typical entries%
	\[
	\frac{Z_{k}^{\prime}\partial_{Z_{k}} G_{j}B_{i}\left(  \alpha|Z\right)}{\alpha(1-\alpha)}, \quad j,k\in\mathcal{I}\setminus \{i\}
	\]
	is full rank.}
\bigskip

Assumption I is not binding in the private value case due to the condition  $\mathcal{I}\neq \{i\}$. Observe also that, for $\alpha=0$ and $\alpha=1$ respectively, $\partial_{Z_{k}} G_{j}B_{i}\left(  \alpha|Z\right)/\alpha$ and $\partial_{Z_{k}} G_{j}B_{i}\left(  \alpha|Z\right)/(1-\alpha)$ stands for  the limit $\partial_{Z_{k}} g_{j}b_{i}\left(  \alpha|Z\right)$.
Note that a stronger rank condition than in Assumption I is in principle testable using the whole set of bidders
$\{1,\ldots,n\}$ instead of the unknown $\mathcal{I}$, as 
$Z_{k}^{\prime}\partial_{Z_{k}} G_{j}B_{i}\left(  \alpha|Z\right)  $ can be consistently estimated. 

The
important role played by $G_{j}B_{i}\left(  \alpha|Z\right)  $ is better
illustrated looking at Lemma \ref{Bidder2} below, which shows in particular that the
increasing strategy $s_{i}\left(  \cdot|Z\right)  $ is identical to the bid
quantile function $B_{i}\left(  \cdot|Z\right)  $. If the other bidders $j$
use strictly increasing and continuous strategies $s_{j}\left(  \cdot
|Z\right)  $, then%
\[
G_{j}B_{i}\left(  \alpha|Z\right)  =s_{j}^{-1}\left[  s_{i}\left(
\alpha|Z\right)  \right]
\]
which is an indicator of asymmetric bidding. The rank condition in Assumption
I fails in particular if $G_{j}B_{i}\left(  \alpha|Z\right)  =\alpha$ for
all $j$, which means that the buyers bid using the same symmetric strategy
$s\left(  \alpha|Z\right)  $. In such case, the variable $Z$ plays a role
similar to a characteristic of the auctioned good. If the bids are drawn from
a Bayesian Nash equilibrium, the non-identification argument of Laffont and
Vuong (1996) holds, showing that there is a private information value model
which is observationally equivalent to the one at hand. The rank condition does not hold if
$G_{j}B_{i}\left(  \alpha|Z\right)  =C_{ji}\left(  \alpha\right)  $ is independent of $Z$, which gives
that $s_{j}\left(  \alpha|Z\right)  =s_{i}\left[  C_{ji}^{-1}\left(  \alpha\right)
|Z\right]  $ for all $j\neq i$ in $\mathcal{I}$.

\section{Main identification results}

\setcounter{equation}{0}

This section considers first identification of the signal distribution.
Section \ref{Pirvf} then derives the identification implications for the
valuation function of the best response condition (\ref{Brc}) in Assumption
S-(i). In particular, the revisited Wilson model is shown to be identified.
However, other valuation function models may not be identified for all signal
values. As shown in Section \ref{Ivf}, the mixed signal value function is
identified in full generality, which ensures that this specification can be
used for counterfactuals that requests to know value functions for all
possible signals. The latter includes in particular the computation of an
expected revenue, which involves integration over all possible signals.

\subsection{Bidding strategy and signal distribution}

The next lemma directly follows from the increasing strategy assumption. Lemma
\ref{Bidder2}-(ii) shows that the bid quantile function is the bid strategy
function, while identification of the distribution of the signal vector $A$
given $Z$ follows from (i). Lemma \ref{Bidder2}-(iii) will be used for identifying the valuation function
later on.

\begin{lemma}
\label{Bidder2} Suppose Assumptions $\ $A and S-(ii,iii,iv) hold. Then

\begin{enumerate}
\item \textbf{[Signal identification]}\textit{ For each} $j=1,\ldots,n$,
\textit{the signals }$A_{j}$ satisfy%
\[
A_{j}=G_{j}\left(  B_{j}|Z\right)  \text{.}%
\]
and are therefore identified, as the conditional signal distribution.

\item \textbf{[Signal bid function identification] }\textit{For each}
$j=1,\ldots,n$, the signal bid function satisfies%
\[
s_{j}\left(  \alpha;Z\right)  =B_{j}\left(  \alpha|Z\right)  \text{.}%
\]

\item \textbf{[Winning probability identification] }Suppose bidder $i$ bid is
$s_{i}\left(  a;Z\right)  $ while her signal $A_{i}$ is equal to $\alpha$.
Then the probability $\omega_{i}\left(  a|\alpha,Z\right)  $ that bidder $i$
wins the auction given $A_{i}=\alpha$ and $Z$ is identified and is equal to%
\[
\omega_{i}\left(  a|\alpha,Z\right)  =\mathbb{P}\left[  \left.  B_{i}\left(
a|Z\right)  >\max_{1\leq j\neq i\leq n}B_{j}\right\vert A_{i}=\alpha,Z\right]
\]

\end{enumerate}
\end{lemma}

\textbf{Proof of Lemma \ref{Bidder2}}: see Appendix.

\subsection{A preliminary identification result for the valuation function
\label{Pirvf}}

The next Lemma is an asymmetric version of first-order condition that
determines the bidding strategy in Milgrom and Weber (1982), see also Laffont
and Vuong (1996), Guerre, Perrigne and Vuong (2000) and Haile et al. (2003)
for econometric applications.
A distinctive feature of the quantile approach
developed here comes from Lemma \ref{Bidder2}-(ii), which shows that the
bidding strategy $s_{i}\left(  \cdot;Z\right)  $ is equal to the bid quantile
$B_{i}\left(  \cdot|Z\right)  $ because $s_{i}\left(  \cdot;Z\right)  $ is
strictly increasing and continuous by Assumption S-(iv). It follows that the best response condition (\ref{Brc}) is equivalent to%
\begin{equation}
\alpha=\arg\max_{a\in\left[  0,1\right]  }\mathbb{E}\left[  \left(
V_{i}\left(  A;Z\right)  -B_{i}\left(  a|Z\right)  \right)  \mathbb{I}\left\{
B_{i}\left(  a|Z\right)  \geq\max_{1\leq j\neq i\leq n}B_{j}\right\}
\left\vert A_{i}=\alpha,Z\right.  \right]  \label{Brca}%
\end{equation}
for all $\alpha$ in $\left[  0,1\right]  $ under Assumption S. Define%
\begin{equation}
\overline{V}_{i}\left(  a|\alpha,Z\right)  =\mathbb{E}\left[  V_{i}\left(
A;Z\right)  \mathbb{I}\left\{  B_{i}\left(  a|Z\right)  \geq\max_{1\leq j\neq
i\leq n}B_{j}\right\}  \left\vert A_{i}=\alpha,Z\right.  \right]  .
\label{BarU}%
\end{equation}
Observe that the expected payoff in (\ref{Brca}) is equal to%
\[
\overline{V}_{i}\left(  a|\alpha,Z\right)  -B_{i}\left(  a|Z\right)
\omega_{i}\left(  a|\alpha,Z\right)
\]
and that%
\[
\left\{  B_{i}\left(  a|Z\right)  \geq\max_{1\leq j\neq i\leq n}B_{j}\right\}
=\bigcup\limits_{1\leq j\neq i\leq n}\left\{  A_{j}\leq G_{j}\left[  \left.
B_{i}\left(  a|Z\right)  \right\vert Z\right]  \right\}  .
\]
Since $B_{i}\left(  \cdot|Z\right)  =s_{i}\left(  \cdot|Z\right)  $,
$G_{j}\left(  \cdot|Z\right)  =s_{j}^{-1}\left(  \cdot|Z\right)  $ and because the p.d.f
$c\left(  \cdot|Z\right)  $ are continuously differentiable by Assumptions
S-(iv) and A respectively, so are $\overline{V}_{i}\left(  \cdot
|\alpha,Z\right)  $ and $\omega_{i}\left(  \cdot|\alpha,Z\right)  $. The
first-order condition associated with (\ref{Brca}) therefore implies%
\begin{equation}
\left.  \frac{\partial\overline{V}_{i}\left(  a|\alpha,Z\right)  }{\partial
a}\right\vert _{a=\alpha}-B_{i}\left(  \alpha;Z\right)  \left.  \frac
{\partial\omega_{i}\left(  a|\alpha,Z\right)  }{\partial a}\right\vert
_{a=\alpha}-B_{i}^{\left(  1\right)  }\left(  \alpha;Z\right)  \omega
_{i}\left(  \alpha|\alpha,Z\right)  =0. \label{FOC2}%
\end{equation}
Define%
\[
\Omega_{i}\left(  \alpha|Z\right)  =\frac{\omega_{i}\left(  \alpha
|\alpha,Z\right)  }{\left.  \frac{\partial\omega_{i}\left(  a|\alpha,Z\right)
}{\partial a}\right\vert _{a=\alpha}},\quad U_{i}\left(  \alpha|Z\right)
=\frac{\left.  \frac{\partial\overline{V}_{i}\left(  a|\alpha,Z\right)
}{\partial a}\right\vert _{a=\alpha}}{\left.  \frac{\partial\omega_{i}\left(
a|\alpha,Z\right)  }{\partial a}\right\vert _{a=\alpha}}.
\]
Rearranging (\ref{FOC2}) gives the next Lemma.
\begin{lemma}
\label{Ident2} Under Assumptions A and S, it holds for each $Z$ of
$\mathcal{Z}$ and all $\alpha$ in $\left[  0,1\right]  $%
\begin{equation}
U_{i}\left(  \alpha|Z\right)  =B_{i}\left(  \alpha|Z\right)  +B_{i}^{\left(
1\right)  }\left(  \alpha|Z\right)  \Omega_{i}\left(  \alpha|Z\right)  
\label{B2U}%
\end{equation}
and $U_{i}\left(  \cdot | \cdot \right)$ is identified.
\end{lemma}
As $\Omega_{i}\left(  \cdot|Z\right)  $ is identified by Lemma \ref{Bidder2}%
-(iii), Equation (\ref{B2U}) in Lemma \ref{Ident2} shows that $U_{i}\left(
\cdot|Z\right)  $ is identified. The merits and limitations of this
identification result are now discussed for the two and three bidders general
case and for the revisited Wilson model.

\subsubsection{Two bidders general case}

It is assumed here that the bidder covariate are of dimension 1 and that
$V_{i}\left(  A;Z\right)  $ is a general valuation function. Suppose without
loss of generality that $i=1$. Observe that the p.d.f of $A_{2}$ given
$A_{1}=\alpha$ and $Z$ is $c\left(  \alpha,\cdot|Z\right)  $ as $A_{1}$ has a
uniform distribution over $\left[  0,1\right]  $ given $Z$. Recall that
$G_{2}B_{1}\left(  a|Z\right)  =G_{2}\left[  B_{1}\left(  a|Z\right)
|Z\right]  $ has a positive derivative $g_{2}b_{1}\left(  a|Z\right)  $ by
Lemma \ref{Bidder2}-(ii) and Assumption S-(iii). This gives for $\overline
{V}_{1}\left(  a|\alpha,Z\right)  $ as in (\ref{BarU}) and $\omega_{1}\left(
a|\alpha,Z\right)  $ as in Lemma \ref{Bidder2}-(iii)
\begin{align*}
\overline{V}_{1}\left(  a|\alpha,Z\right)   &  =\int_{0}^{1}V_{1}\left(
\alpha,t;Z\right)  \mathbb{I}\left[  B_{1}\left(  a|Z\right)  \geq
B_{2}\left(  t|Z\right)  \right]  c\left(  \alpha,t|Z\right)  dt\\
&  =\int_{0}^{G_{2}B_{1}\left(  a|Z\right)  }V_{1}\left(  \alpha,t;Z\right)
c\left(  \alpha,t|Z\right)  dt,\\
\omega_{1}\left(  a|\alpha,Z\right)   &  =\int_{0}^{1}\mathbb{I}\left[
B_{1}\left(  a|Z\right)  \geq B_{2}\left(  t|Z\right)  dt\right]  c\left(
\alpha,t|Z\right)  dt=\int_{0}^{G_{2}B_{1}\left(  a|Z\right)  }c\left(
\alpha,t|Z\right)  dt
\end{align*}
so that%
\begin{align*}
\frac{\partial\overline{V}_{1}\left(  a|\alpha,Z\right)  }{\partial a} &
=g_{2}b_{1}\left(  a|Z\right)  V_{1}\left(  \alpha,G_{2}B_{1}\left(
a|Z\right)  ;Z\right)  c\left(  \alpha,G_{2}B_{1}\left(  a|Z\right)
|Z\right)  \\
\frac{\partial\omega_{1}\left(  a|\alpha,Z\right)  }{\partial a} &
=g_{2}b_{1}\left(  a|Z\right)  c\left(  \alpha,G_{2}B_{1}\left(  a|Z\right)
|Z\right)  .
\end{align*}
Hence%
\begin{equation}
U_{1}\left(  \alpha|Z\right)  =\frac{\left.  \frac{\partial\overline{V}%
_{1}\left(  a|\alpha,Z\right)  }{\partial a}\right\vert _{a=\alpha}}{\left.
\frac{\partial\omega_{1}\left(  a|\alpha,Z\right)  }{\partial a}\right\vert
_{a=\alpha}}=V_{1}\left(  \alpha,G_{2}B_{1}\left(  \alpha|Z\right)  ;Z\right)
.\label{U1I2}%
\end{equation}

It follows from Lemma \ref{Ident2} that, for each $Z$, $V_{1}\left(
\alpha_{1},\alpha_{2};Z\right)  $ is nonparametrically identified over the
curve
\[
\left\{  \left(  \alpha_{1},\alpha_{2}\right)  ;\alpha_{2}=G_{2}B_{1}\left(
\alpha_{1}|Z\right)  ,\alpha_{1}\in\left[  0,1\right]  \right\}  .
\]
This is insufficient to identify $V_{1}\left(  \cdot\right)  $ over $\left[
0,1\right]  ^{2}\times\mathcal{Z}$ nonparametrically. However identification
may hold under further restrictions of the valuation function as detailed here.

\bigskip

\begin{itemize}
\item \textbf{Bidder covariate exclusion restriction.} Somaini (2018)
considers the exclusion restriction $V_{1}\left(  \alpha_{1},\alpha
_{2};Z\right)  =V_{1}\left(  \alpha_{1},\alpha_{2};Z_{1}\right)  $, which, for
a given $Z_{1}$, ensures identification of the latter for any $\alpha_{1}$ in
$\left[  0,1\right]  $ and $\alpha_{2}$ in $\left[  \min_{Z_{2}}G_{2}%
B_{1}\left(  \alpha_{1}|Z\right)  ,\max_{Z_{2}}G_{2}B_{1}\left(  \alpha
_{1}|Z\right)  \right]  $ by a proper choice of $Z_{2}$.

\item \textbf{Separability restriction. }Consider the additive specification
\[
V_{1}\left(  \alpha_{1},\alpha_{2};Z\right)  =v_{1}\left(  \alpha_{1}%
,\alpha_{2}\right)  +v_{0}\left(  Z\right)
\]
with the normalization $v_{1}\left(  0,0\right)  =0$.\footnote{As discussed for the mixed signal specification, the common initial strategy condition of Assumption S-(ii) holds for Bayesian Nash equilibrium bids if all bidders have the same $v_0(Z)$. Assuming an exclusion restriction $v_{0i} (Z) = v_{0i} (Z_i)$ as in Somaini (2018) is also possible, using for identification purpose values $Z$ satisfying 
$v_{01}(Z_1)=\cdots v_{0n} (Z_n)$, which are identified by the participation of all bidders. It is also desirable to assume that $A$ and $Z$ are independent as identifying the conditional distribution of $A$ given $Z$ is difficult when S-(ii) does not hold.} As $G_{2}B_{1}\left(
0|Z\right)  =0$ under Assumption S-(ii), $v_{2}\left(  Z\right)  =U_{1}\left(
0|Z\right)  $ is identified and so is $v_{1}\left(  \alpha_{1},G_{2}%
B_{1}\left(  \alpha_{1}|Z\right)  \right)  =U_{1}\left(  \alpha_{1}|Z\right)
-U_{1}\left(  0|Z\right)  $. It then follows that the valuation function of
bidder 1 is identified for any $\alpha_{1}$ in $\left[  0,1\right]  $ and
$\alpha_{2}$ in $\left[  \min_{Z}G_{2}B_{1}\left(  \alpha_{1}|Z\right)
,\max_{Z}G_{2}B_{1}\left(  \alpha_{1}|Z\right)  \right]  $, which may differ from $[0,1]$. This restriction on $\alpha_2$ can be removed under an additional additive assumption. Suppose now $v_1 (\alpha_1,\alpha_2)=v_{11} (\alpha_1)+v_{12} (\alpha_2)$ with $v_{11} (0)=v_{12} (0)=0$. Then $U_1(\cdot|Z)$ identifies
\[
u_1 (\alpha|Z)=v_{11} (\alpha) + v_{12} \left(G_{2}B_{1}\left(  \alpha|Z\right)\right).
\]
As $\partial_{Z} u_1 (\alpha|Z) =  v_{12}^{(1)} \left(G_{2}B_{1}\left(  \alpha_{1}|Z\right)\right) \partial_{Z}G_{2}B_{1}\left(  \alpha|Z\right)$, it follows that $v_{12}^{(1)} (\cdot)$ is identified over $[0,1]$ if, for each $\alpha$ in $[0,1]$, there exists a $Z$ such that $\partial_{Z}G_{2}B_{1}\left(  \alpha |Z\right)\neq 0$. Hence the initial condition $v_{12} (0)=0$ yields that $v_{12} (\cdot)$ is identified, and then $v_{11} (\cdot)$ is also identified, both over the whole $[0,1]$.

\item \textbf{Signal exclusion restriction. }The value function is identified
in the private value case $V_{1}\left(  \alpha_{1},\alpha_{2};Z\right)
=V_{1}\left(  \alpha_{1};Z\right)  $. The signal exclusion restriction
$V_{1}\left(  \alpha_{1},\alpha_{2};Z\right)  =V_{1}\left(  \alpha
_{2};Z\right)  $ yields, for each $Z$, identification for all $\alpha_{2}$
between $\min_{\alpha_{1}\in\left[  0,1\right]  }G_{2}B_{1}\left(  \alpha
_{1}|Z\right)  $ and $\max_{\alpha_{1}\in\left[  0,1\right]  }G_{2}%
B_{1}\left(  \alpha_{1}|Z\right)  $, i.e. in $\left[  0,1\right]  $ under
Assumption S-(ii) and the conditions in Lizerri and Persico (2000).
\end{itemize}

Hence identification may not hold for all signals, possibly preventing to implement some
counterfactuals such as computation of an optimal reserve price. As seen from Theorems \ref{MSMidentn2} and \ref{MSMidentn3} below, this contrasts with the
mixed signal value functions considered here.

\subsubsection{Three bidder general case}

The case of a larger number $n$ of bidders is more difficult because the
identified expected value is a multiple integral of order $n-2$. To see this,
suppose that $n=3$ and that the valuation of interest is the one of the first
bidder. The p.d.f of $\left(  A_{2},A_{3}\right)  $ given $A_{1}=\alpha$ and
$Z$ is $c\left(  \alpha,\cdot,\cdot|Z\right)  $ and $\overline{V}_{1}\left(
a|\alpha,Z\right)  $, $\omega_{1}\left(  a|\alpha,Z\right)  $ are now given
by
\begin{align*}
\overline{V}_{1}\left(  a|\alpha,Z\right)   &  =\int V_{1}\left(  \alpha
,t_{2},t_{3};Z\right)  \mathbb{I}\left[  B_{1}\left(  a|Z\right)  \geq
\max\left\{  B_{2}\left(  t_{3}|Z\right)  ,B_{2}\left(  t_{3}|Z\right)
\right\}  \right]  c\left(  \alpha,t_{2},t_{3}|Z\right)  dt_{2}dt_{3}\\
&  =\int_{0}^{G_{3}B_{1}\left(  a|Z\right)  }\left[  \int_{0}^{G_{2}%
B_{1}\left(  a|Z\right)  }V_{1}\left(  \alpha,t_{2},t_{3};Z\right)  c\left(
\alpha,t_{2},t_{3}|Z\right)  dt_{2}\right]  dt_{3},\\
\omega_{1}\left(  a|\alpha,Z\right)   &  =\int_{0}^{G_{3}B_{1}\left(
a|Z\right)  }\left[  \int_{0}^{G_{2}B_{1}\left(  a|Z\right)  }c\left(
\alpha,t_{2},t_{3}|Z\right)  dt_{2}\right]  dt_{3}.
\end{align*}
Hence%
\begin{align}
&  \left.  \frac{\partial\omega_{1}\left(  a|\alpha,Z\right)  }{\partial
a}\right\vert _{a=\alpha}\times U_{1}\left(  \alpha|Z\right)  \nonumber\\
&  \quad\quad\quad=g_{3}b_{1}\left(  \alpha|Z\right)  \int_{0}^{G_{2}%
B_{1}\left(  \alpha|Z\right)  }V_{1}\left(  \alpha,t_{2},G_{3}B_{1}\left(
\alpha|Z\right)  ;Z\right)  c\left(  \alpha,t_{2},G_{3}B_{1}\left(
\alpha|Z\right)  |Z\right)  dt_{2}\nonumber\\
&  \quad\quad\quad+g_{2}b_{1}\left(  \alpha|Z\right)  \int_{0}^{G_{3}%
B_{1}\left(  \alpha|Z\right)  }V_{1}\left(  \alpha,G_{2}B_{1}\left(
\alpha|Z\right)  ,t_{3};Z\right)  c\left(  \alpha,G_{2}B_{1}\left(
\alpha|Z\right)  ,t_{3}|Z\right)  dt_{3},\label{U1I3}%
\end{align}%
\begin{align*}
\left.  \frac{\partial\omega_{1}\left(  a|\alpha,Z\right)  }{\partial
a}\right\vert _{a=\alpha} &  =g_{3}b_{1}\left(  \alpha|Z\right)  \int%
_{0}^{G_{2}B_{1}\left(  \alpha|Z\right)  }c\left(  \alpha,t_{2},G_{3}%
B_{1}\left(  \alpha|Z\right)  |Z\right)  dt_{2}\\
&  +g_{2}b_{1}\left(  \alpha|Z\right)  \int_{0}^{G_{3}B_{1}\left(
\alpha|Z\right)  }c\left(  \alpha,G_{2}B_{1}\left(  \alpha|Z\right)
,t_{3}|Z\right)  dt_{3}.
\end{align*}
Lemma \ref{Ident2} therefore establishes nonparametric identification of an
integral function of the valuation function over a set of signal variables
$\left(  \alpha_{1},\alpha_{2},\alpha_{3}\right)  =\left(  \alpha_{1}%
,G_{2}B_{1}\left(  \alpha_{1}|Z\right)  ,G_{3}B_{1}\left(  \alpha
_{1}|Z\right)  \right)  $. As for the two bidder case, this may not be useful
for most applications without further restriction on the valuation function.
Somaini (2018) derives some identification results under covariate
exclusion.

\subsubsection{Revisited Wilson model}

On the other hand, the revisited Wilson model is an example of specification
that can be easily identified from Lemma \ref{Ident2}. Indeed (\ref{Wilson})
and the expressions of $U_{1}\left(  \alpha|Z\right)  $ derived above imply%
\[
\lim_{Z_{1}\rightarrow+\infty}U_{1}\left(  \alpha|Z\right)  =\gamma_{0}\left(
\alpha\right)
\]
because $Z_{1}=+\infty$ means that bidder 1 is perfectly informed about the
value of the good. This is sufficient to recover identification of this
specification.

\subsection{Identification of the mixed signal valuation function \label{Ivf}}

The mixed signal valuation specification (\ref{MSMU}) can be identified using a three step procedure. 
First an initial or terminal value for the slope functions $\gamma_{ij} (\cdot)$ is identified.
Second, the function $\Phi_{i}\left(
\cdot\right)  $ is identified. Third, thanks to the rank condition I, the slope functions are then determined as
solving a differential, or an integro-differential, equation. Identification of the slope functions holds over the full set $\left[  0,1\right]^{n}$ of signals, as suitable for many counterfactual applications. 
However the identification procedure implementation importantly differs depending whether there are two bidders or more, especially in the first step and in the choice of identifying an initial or a terminal value for the slope functions.

\subsubsection{Two bidder case\label{2bc}}

Suppose $n=2$ and let bidder $1$ $\Phi_1 (\cdot)$ and $(\gamma_{11} (\cdot),\gamma_{12} (\cdot))$ be the parameters of interest. Hence Lemma \ref{Ident2} and (\ref{U1I2}) show that
\begin{equation}
U_{1}\left(  \alpha|Z\right)  =\Phi_{1}\left[  Z_{1}^{\prime}\gamma_{11}\left(
\alpha\right)  ,Z_{2}^{\prime}\gamma_{12}\left[  G_{2}B_{1}\left(  \alpha|Z\right)
\right]  \right]  
\label{MSVident0}
\end{equation}
is identified. The considered next step  is the identification of a terminal value for $(\gamma_{11} (\cdot),\gamma_{12} (\cdot))$.

\paragraph{Step 1: identification of $\mathcal{I}$ and $(\gamma_{11} (1),\gamma_{12} (1))$. }
As $G_{2}B_{1}\left(  1|Z\right)  =1$ for all $Z$ by the terminal condition of Assumption
S-(ii) and Lemma \ref{Bidder2}-(ii), setting $\alpha=1$ in (\ref{MSVident0}) yields the identity
\[
U_{1}\left(  1 |Z\right)  =
\Phi_{1}\left( 
Z_{1}^{\prime}
\gamma_{11}\left(1\right)  
,
Z_{2}^{\prime}
\gamma_{12}  \left(  1\right)
\right)
\]
and 
$
\Phi_{1}\left( 
Z_{1}^{\prime}
\gamma_{11}\left(1\right)  
,
Z_{2}^{\prime}
\gamma_{12}  \left(  1\right)
\right)
$ 
is therefore identified. 
As all $\gamma_{1j} (1)$  are not $0$ under Assumption G-(ii), Assumption P ensures that $j$ does not belong to $\mathcal{I}$ if and only if
\[
\partial_{Z_k}
\left[
U_{1}\left(  1 |Z\right)
\right]
=
\frac{\partial 
\Phi_{1}\left( 
	Z_{1}^{\prime}
	\gamma_{11}\left(1\right)  
	,
	Z_{2}^{\prime}
	\gamma_{12}  \left(  1\right)
\right)}{\partial x_j}
\gamma_{1j}  \left(  1\right)
=
0
\text{ for all $Z$ in $\mathcal{Z}$.}
\]
This implies that 
$\mathcal{I} = \left\{ j; \frac{\partial \Phi_1 (\cdot,\cdot)}{\partial x_j} \neq 0 \right\}$ 
is identified. 
As 
\[
\lim_{Z \rightarrow 0}
\partial_{Z_j} U_{1}\left(  1 | Z\right)  
=
\frac{\partial \Phi_{1} \left( 0,0\right)}{\partial x_j}
\gamma_{1j}\left(1\right)
\]
for a smooth $\Phi_1 (\cdot)$ satisfying Assumption P, using the normalization (\ref{Normphi}) shows that
$\gamma_{1j} (1)$ is identified for all $j$ in $\mathcal{I}$.

\paragraph{Step 2: identification of $\Phi_1 (\cdot)$.} Consider $x=(x_1,x_3)$ in $\mathbb{R}_{+*}^2$. For $k$ in the identified $\mathcal{I}$, there is a $Z_k$ in $\mathbb{R}_{+*}^D$ such that $x_k = Z_k^{\prime} \gamma_{1k} (1)$, recalling that $\gamma_{1k} (1)$ has been identified in the preceding step. When $k$ does not belong to $\mathcal{I}$, choose an arbitrary $Z_k$ in $\mathbb{R}_{+*}^D$. It then holds for such choice of $Z_1$ and $Z_2$
\[
\Phi_1 (x_1,x_2)
=
\Phi_1 
\left(
Z_1^{\prime} \gamma_{11} (1),Z_2^{\prime} \gamma_{12} (1)
\right)
\]
so that $\Phi_1(\cdot)$ is identified over $\mathbb{R}_{+*}^2$. Under Assumption P, continuity of $\Phi_1(\cdot)$ ensures it is identified over $\mathbb{R}_{+}^2$.

\paragraph{Step 3: identification of $(\gamma_{11} (\cdot),\gamma_{12} (\cdot))$. }
Suppose first the private value case $\mathcal{I}=\{1\}$, ie $\Phi_1 (x_1,x_2)=\Phi_1 (x_1)$. Then monotonicity in Assumption P and
(\ref{MSVident0}) show that $Z_1^{\prime} \gamma_1 (\cdot) =\Phi_1^{-1} \left[U(\cdot|Z_1,Z_2)\right]$ for any $(Z_1,Z_2)$ in $\mathcal{Z}$. If $\mathcal{I}=\{2\}$, ie bidder 1 is uninformed, $\gamma_2 \left[G_2 B_1 (\alpha|Z)\right]$ is similarly identified, which ensures that $\gamma_2 (\cdot)$ as the identified $G_2 B_1 (\cdot|Z)$ is one to one by Lemma \ref{Bidder2}-(ii) under Assumption S-(iv).

Consider now the case where $\mathcal{I=}\left\{  1,2\right\}$. Observe that, since $g_{2}b_{1}\left(  \cdot|Z\right)  >0$ by Assumption S-(iv) and Lemma \ref{Bidder2}-(ii),
\begin{align}
	\partial_{Z_2}\left\{  \gamma_{12}\left[  G_{2}B_{1}\left(  \alpha|Z\right)
		\right]  \right\}   
	&  =\gamma_{12}^{\left(  1\right) }\left[  G_{2}B_{1}\left(  \alpha|Z\right)  \right]  
	\partial_{Z_2} G_{2} B_{1}\left(  \alpha|Z\right)  \nonumber\\
	&  =\gamma_{12}^{\left(  1\right)  }\left[  G_{2}B_{1}\left(  \alpha|Z\right)
	\right]  g_{2}b_{1}\left(  \alpha|Z\right)  
	\frac{\partial_{Z_2} G_{2} B_{1}\left(  \alpha|Z\right)}{g_{2}b_{1}\left(
		\alpha|Z\right)  }\nonumber\\
	&  =\frac{\partial\left\{  \gamma_{12}\left[  G_{2}B_{1}\left(  \alpha
		|Z\right)  \right]  \right\}  }{\partial\alpha}
	\frac{\partial_{Z_2} G_{2} B_{1}\left(  \alpha|Z\right)}{g_{2}b_{1}\left(
		\alpha|Z\right)  }.\label{Dz2da}%
	\end{align}
Differentiating $U_{1}\left(  \alpha|Z\right)  $ with respect to $\alpha$ and $Z_{2}$ then gives%
\begin{align*}
	&  \frac{\partial\Phi_{1}}{\partial x_{1}}\left[  \gamma_{11}\left(
	\alpha\right)  Z_{1},\gamma_{12}\left[  G_{2}B_{1}\left(  \alpha|Z\right)
	\right]  Z_{2}\right]  Z_{1}\frac{d\gamma_{11}\left(  \alpha\right)  }%
	{d\alpha}\\
	&  \quad+\frac{\partial\Phi_{1}}{\partial x_{2}}\left[  \gamma_{11}\left(
	\alpha\right)  Z_{1},\gamma_{12}\left[  G_{2}B_{1}\left(  \alpha|Z\right)
	\right]  Z_{2}\right]  Z_{2}\frac{\partial\left\{  \gamma_{12}\left[
		G_{2}B_{1}\left(  \alpha|Z\right)  \right]  \right\}  }{\partial\alpha}%
	=\frac{\partial U_{1}\left(  \alpha|Z\right)  }{\partial\alpha},
\end{align*}%
\begin{align*}
	&  \frac{\partial\Phi_{1}}{\partial x_{2}}\left[  \gamma_{11}\left(
	\alpha\right)  Z_{1},\gamma_{12}\left[  G_{2}B_{1}\left(  \alpha|Z\right)
	\right]  Z_{2}\right]  \gamma_{12}\left[  G_{2}B_{1}\left(  \alpha|Z\right)
	\right]  \\
	&  \quad+\frac{\partial\Phi_{1}}{\partial x_{2}}\left[  \gamma_{11}\left(
	\alpha\right)  Z_{1},\gamma_{12}\left[  G_{2}B_{1}\left(  \alpha|Z\right)
	\right]  Z_{2}\right]  Z_{2}\frac{\partial_{Z_2} G_{2}B_{1}\left(
			\alpha|Z\right)  }{g_{2}b_{1}\left(  \alpha|Z\right)  }%
	\frac{\partial\left\{  \gamma_{12}\left[  G_{2}B_{1}\left(  \alpha|Z\right)
		\right]  \right\}  }{\partial\alpha}=\partial_{Z_2} U_{1}\left(
	\alpha|Z\right)
\end{align*}
which shows that $\Gamma_{1}\left(  \alpha|Z\right)  =\left[  \gamma_{11}\left(  \alpha\right)  ,\gamma_{12}\left[  G_{2}B_{1}\left(\alpha|Z\right)  \right]  \right]^{\prime}$ are the solution of a $2\times2$
system of differential equations with the terminal condition $\Gamma_{1}\left(  1|Z\right) = [1,1]^{\prime}$. 
Standard uniqueness of the solution of such differential systems would then ensure that $\Gamma_{1}\left(  \cdot|Z\right)$, and then $[\gamma_{11} (\cdot),\gamma_{12} (\cdot) ]$, is identified. Unfortunately, this argument cannot be applied here because the item  $\partial_{Z_2} G_{2}B_{1}\left(
\alpha|Z\right)$, which appears in front of $\gamma_{11}^{(1)} (\cdot)$ in the differential system, vanishes when $\alpha=1,0$ due to the terminal and initial bidding strategy conditions in Assumption S-(ii,iii) and Lemma \ref{Bidder2}, which identifies strategy and bid quantile functions. This issue is addressed in the proof of the next Theorem in the Appendix.\footnote{
As noted by a Referee, taking $Z_2$ and $Z_1$ equal to $0$  would give, respectively 
\begin{eqnarray*}
Z_{1}^{\prime} \gamma_{11} (\alpha) 
& = & 
 \Phi_1^{-1|x_1} \left[U_1 (\alpha|Z_1,0),0 \right],
\\
Z_{2}^{\prime} \gamma_{12} (\alpha)
& = &
\Phi_1^{-1|x_2} \left[0,U_1 ( (G_2B_1(\alpha|0,Z_2))|0,Z_2) \right]
,
\end{eqnarray*}
where $\Phi_1^{-1|x_1} (y,\cdot)$ is the inverse of $x_2 \mapsto \Phi_1 (y,x_2)$ and $\Phi_1^{-1|x_2} (\cdot,y)$ is similarly defined. Hence $\gamma_{11} (\cdot)$ and $\gamma_{12} (\cdot)$ would be identified. 
This may be however difficult to implement for Bayesian Nash Equilibrium bids. For instance, in the private value case with bidder 2 value $Z_2^{\prime} \gamma_{22} (A_2)$,  the bidding strategies $B_1 (\cdot|Z)$ and $B_2(\cdot|Z)$ may have degenerate limits, if any, when $Z_2$ goes to $0$, because the fact that bidder 2 value goes to 0 is also known to bidder 1. As shown by (\ref{MSVident0}), it follows that establishing the existence of
$U_1 (\alpha|Z_1,0)$, $U_1 ( (G_2B_1(\alpha|0,Z_2))|0,Z_2)$, or even $\lim_{Z_{1} \downarrow 0} U_1 (\alpha|Z)$, $\lim_{Z_{2} \downarrow 0}U_1 ( (G_2B_1(\alpha|Z))|Z)$, may be difficult. \label{gb} }
Theorem \ref{MSMidentn2} summarizes the identification result.

\begin{theorem}
	\label{MSMidentn2}
	Suppose Assumptions A, I, G with the terminal condition G-(i), P and Z hold and that $n=2$. Then, up to a normalization of $\Phi_i(\cdot,\cdot)$ and $(\gamma_{11} (\cdot),\gamma_{12} (\cdot))$ as (\ref{Normphi}): 
	\begin{enumerate}
		\item The set $\mathcal{I}$ of  active signals defined in (\ref{Phisupp}) is identified;
		\item The function $\Phi_{i}\left(  x_{1},x_2\right)  $ is identified
		over $\mathbb{R}^2_{+}$;
		\item The slope functions $\gamma_{ij}\left(  \cdot\right)  $ are identified
		for all $j$ in $\mathcal{I}$.
	\end{enumerate}
\end{theorem}

\textbf{Proof of Theorem \ref{MSMidentn2}}: see Appendix.

\bigskip
Note that identification of the slope makes use of values of $Z$ in the vicinity of $0$, plus an arbitrary non vanishing $Z$, so that the slopes are overidentified.
This identification procedure also works when bidder $i$ makes dominated bids below a threshold 
$\underline{\alpha}_i (Z)$, which is identified by the probability of observing such bids, as it is sufficient to solve the differential system over $[\underline{\alpha}_i (Z),1 ]$. Identification of the slope would then hold over $[\inf_{Z} \underline{\alpha}_i (Z),1 ]$.

In the two bidder case, identification is based upon the terminal strategy condition S-(iii), which is quite satisfactory here as it is based on a well established result of Lizzeri and Persico (2000) when bids are drawn from a Bayesian Nash Equilibrium with valuation functions from the mixed signal specification. It implies that the initial slope values are unconstrained by Assumption G. Hence both conditions BNE-(i) and (ii) can be used to ensure that the initial strategy condition S-(ii) holds with Bayesian Nash Equilibrium bids.

\subsubsection{More than two bidders \label{3bc0}}

We first explain why using, as in the two bidder case, the terminal value of $U_1 (\cdot|Z)$ for identification purpose becomes difficult when the number of bidders is larger.
For the sake of discussion brevity, assume $n=3$, the case of a higher number of bidders being similar, and take $i=1$. 
Let
$
W_1 \left(  \alpha|Z\right)  =\left.  \frac{\partial\overline{V}_{1}\left(
	a|\alpha,Z\right)  }{\partial a}\right\vert _{a=\alpha}%
$ be the function in (\ref{U1I3}), so that $U_1(\alpha|Z) = W_1 \left(  \alpha|Z\right)/ \left(\left.  \frac{\partial\omega_{1}\left(a|\alpha,Z\right)  }{\partial a}\right\vert _{a=\alpha} \right)$. For $\alpha=1$, it holds under the terminal strategy condition in Assumption S-(ii)
\begin{align*}
W_1 \left(  1|Z\right) = & 
g_{3}b_{1}\left(  1|Z\right)  \int_{0}^{1}
\Phi_{1}
\left(  
Z_1^{\prime}
\gamma_{11} (1),
Z_2^{\prime}
\gamma_{12}
\left( t_2\right)
,
Z_3^{\prime}
\gamma_{13} (1)
\right) 
c\left(  t_{2}  |A_1=1,A_3=1, Z\right)  dt_{2}\nonumber\\
&  
+g_{2}b_{1}\left(  1|Z\right)  
\int_{0}^{1}
\Phi_{1}\left(  
Z_1^{\prime} \gamma_{11} (1),
Z_2^{\prime}
\gamma_{12} (1)
,
Z_3^{\prime}
\gamma_{13}
(t_3)
\right)  
c\left(t_3  |A_1=1,A_2=1,Z\right)
dt_{3}.
\end{align*}
Hence $W_1 (1|Z)$ now depends upon the whole slope functions $\gamma_{12} (\cdot)$ and
$\gamma_{13} (\cdot)$, while only the slope terminal values were involved in the two bidder case. Using the terminal value of $U_1 (\cdot|Z)$ for identification purpose does not seem feasible and we use instead the common initial strategy condition of 
Assumption S-(ii).\footnote{Letting $Z$ goes to $0$ allows to identify
$\left(\lim_{Z\downarrow 0} g_{3}b_{1}\left(  1|Z\right) \right) \int_0^1 \gamma_{12} (t) c\left(  t  |A_1=1,A_3=1, 0\right)  dt
+
\left(\lim_{Z\downarrow 0} g_{2}b_{1}\left(  1|Z\right) \right) \gamma_{12} (1)
$
and a similar functional of $\gamma_{13} (\cdot)$, which can be used instead of the first order condition. We did not attempt to implement this approach, which is not straightforward, due to the fact that the limits of $g_j b_1 (1|z)$, $j=2,3$, when $Z$ goes to $0$ may not be well defined as discussed in footnote \ref{gb} for the limits of $G_j B_1 (1|z)$. }

It is first shown in the proof of Theorem \ref{MSMidentn3} that
\begin{equation}
U_{i}\left(  0|Z\right)  =\Phi_{i}\left[  Z_{1}^{\prime}\gamma_{i1}\left(
0\right)  ,\ldots,Z_{n}^{\prime}\gamma_{in}\left(  0\right)  \right],
\label{Ui0}%
\end{equation}
which is identified by Equation (\ref{B2U}) in Lemma \ref{Ident2}.
When all the $\gamma_{ij} (0)$, $j$ in $\mathcal{I}$, differ from $0$ as assumed in Assumption G-(ii), arguing as in the Step 1 of the two bidder case permits to identify $\mathcal{I}$ and those $\gamma_{ij} (0)$. Repeating Step 2 of the two bidder case then yields that $\Phi_i (\cdot)$ is identified.

Identifying the slope functions $\gamma_{ij} (\cdot)$ is slightly more complicated than in Step 3 of the two bidder case. Differentiating the identified $U_i (\alpha|Z)$ with respect to the signal shifters $Z_j$, $j\neq i$ now gives an integro-differential system. 
The proof of Theorem \ref{MSMidentn3}, which summarizes our identification result for $n\geq 3$, establishes uniqueness of its solution.
\begin{theorem}
	\label{MSMidentn3} 
	Suppose Assumptions A, I, G with the initial condition G-(ii), P and Z hold. Then, up to a normalization of $\Phi_i(\cdot,\cdots,\cdot)$ and $(\gamma_{11} (\cdot),\ldots,\gamma_{1n} (\cdot))$ as (\ref{Normphi}): 
	\begin{enumerate}
		\item The set $\mathcal{I}$ of  active signals defined in (\ref{Phisupp}) is identified;
		\item The function $\Phi_{i}\left(  x_{1},\ldots,x_n\right)  $ is identified
		over $\mathbb{R}^n_{+}$;
		\item The slope functions $\gamma_{ij}\left(  \cdot\right)  $ are identified
		for all $j$ in $\mathcal{I}$.
	\end{enumerate}
\end{theorem}

\textbf{Proof of Theorem \ref{MSMidentn3}}: see Appendix.

\bigskip
Theorem \ref{MSMidentn3} also applies to the two bidder case, but relies more importantly on the initial strategy condition in Assumption S-(ii), which is less natural than the terminal one S-(iii) derived in Lizzeri and Persico (2000). Because identification relies on the initial value $U_i (0|Z)$ instead of $U_i (1|Z)$ used for the two bidder case, Theorem \ref{MSMidentn3} makes use of the initial slope Assumption G-(ii), which imposes that $\gamma_{ij} (0)$ must differ from $0$ for all $j$ of $\mathcal{I}$. This has important consequences for Bayesian Nash Equilibrium bids as only the common $\Phi_i (\cdot)$ and initial slope restrictions of Condition BNE-(ii) can be used.\footnote{
A conjecture that would allow to use Condition BNE-(i), which allows for more asymmetric functions $\Phi_i (\cdot)$, is that under Assumption G-(i) which states that $\gamma_{ij} (0)=0$ for all $i,j$, all the strategies $s_i (\tau_i \alpha|Z/\alpha)$ converge when $\alpha$ goes to $0$ because the corresponding valuation functions satisfy
\[
\lim_{\alpha\downarrow 0}
\Phi_i
\left(
\frac{Z_1^{\prime} \gamma_{i1} (\tau_1 \alpha)}{\alpha}
, \ldots ,
\frac{Z_n^{\prime} \gamma_{in} (\tau_n \alpha)}{\alpha}
\right)
=
\Phi_i
\left(
Z_1^{\prime} \gamma_{i1}^{(1)} (0) \tau_1 
, \ldots ,
Z_n^{\prime} \gamma_{in}^{(1)} (0)  \tau_n 
\right).
\]
Setting the $\tau_i$ to $1$ will then allows to identify $\Phi_i (\cdot)$ as the $\gamma_{i1}^{(1)} (0)$ do not vanish and can be identified using $Z \rightarrow 0$ and (\ref{Normphi}). Establishing this conjecture is however out of the scope of the present paper.
}
 This reduces the degrees of bidder asymmetry permitted by the mixed signal specification, as the only possible cause of asymmetry is now given by distinct slopes $\gamma_{ij} (\alpha)$ for $\alpha>0$. As many studies assume that 
$\Phi_i (x_1,\ldots,x_n) = x_1+\cdots+x_n$ as in Somaini (2018) and because the slopes $\gamma_{ij} (\cdot)$ can be very small, this may  nevertheless be flexible enough for many applications.

\setcounter{equation}{0}

\section{Estimation strategy}

The identification proof is constructive and can be directly used for
estimation, although more suitable procedures can be proposed, as a 
procedure similar to the one introduced by Botosaru (2019) for a duration
model with unobserved heterogeneity. Consider for the sake of brevity the two
bidder case. The first stage consists in an estimation of $U_{1}\left(
\alpha|Z\right)  $ based on (\ref{B2U}) in Lemma \ref{Ident2}%
\[
\widehat{U}_{1}\left(  \alpha|Z\right)  =\widehat{B}_{1}\left(  \alpha
|Z\right)  +\widehat{B}_{1}^{\left(  1\right)  }\left(  \alpha|Z\right)
\widehat{\Omega}_{1}\left(  \alpha|Z\right)
\]
where the quantile derivative estimator can be obtained using Gimenes and
Guerre (2019). Recall now that $U_{1}\left(  \alpha|Z\right)  =\Phi_{1}\left[
\gamma_{11}\left(  \alpha\right)  Z_{1},\gamma_{12}\left[  G_{2}B_{1}\left(
\alpha|Z\right)  \right]  Z_{2}\right]  $ by (\ref{MSVident0}), which could be
estimated using
\[
\Phi_{1}\left[  \gamma_{11}\left(  \alpha\right)  Z_{1},\gamma_{12}\left[
\widehat{G}_{2}\widehat{B}_{1}\left(  \alpha|Z\right)  \right]  Z_{2}\right]
.
\]
The second stage of the procedure matches the above with $\widehat{U}%
_{1}\left(  \alpha|Z\right)  $ to produce an estimator of $\Phi_{1}\left(
\cdot\right)  $, $\gamma_{11}\left(  \cdot\right)  $ and $\gamma_{12}\left(
\cdot\right)  $%
\begin{align*}
&  \left[  \widehat{\Phi}_{1}\left(  \cdot\right)  ,\widehat{\gamma}%
_{11}\left(  \cdot\right)  ,\widehat{\gamma}_{12}\left(  \cdot\right)  \right]
\\
&  \quad=\arg\min_{\Phi,\gamma_{1},\gamma_{2}}\int\left\{  \int_{0}^{1}\left(
\widehat{U}_{1}\left(  \alpha|Z\right)  -\Phi\left[  \gamma_{1}\left(
\alpha\right)  Z_{1},\gamma_{2}\left[  \widehat{G}_{2}\widehat{B}_{1}\left(
\alpha|Z\right)  \right]  Z_{2}\right]  \right)  ^{2}d\alpha\right\}  dZ
\end{align*}
where the minimization is performed over a sieve for $\gamma_1 (\cdot)$, $\gamma_2 (\cdot)$ and $\Phi (\cdot)$, or over a simpler set of functions $\Phi(\cdot)$, such as the additive of maximum functions of the two first examples. Up to the estimation
$\widehat{G}_{2}\widehat{B}_{1}\left(  \cdot|\cdot\right)  $, these estimators
can be studied as in Botosaru (2019).

Practical computation of these estimators can be done in the following
iterative way
\[
\widehat{\Phi}_{1,k+1}\left(  \cdot\right)  =\arg\min_{\Phi_{1}}\int\left\{
\int_{0}^{1}\left(  \widehat{U}_{1}\left(  \alpha|Z\right)  -\Phi_{1}\left[
\widehat{\gamma}_{11,k}\left(  \alpha\right)  Z_{1},\widehat{\gamma}%
_{12,k}\left[  \widehat{G}_{2}\widehat{B}_{1}\left(  \alpha|Z\right)  \right]
Z_{2}\right]  \right)  ^{2}d\alpha\right\}  dZ,
\]%
\begin{align}
&  \left[  \widehat{\gamma}_{11,k+1}\left(  \cdot\right)  ,\widehat{\gamma
}_{12,k+1}\left(  \cdot\right)  \right]  \nonumber\\
&  \quad=\arg\min_{,\gamma_{1},\gamma_{2}}\int\left\{  \int_{0}^{1}\left(
\widehat{U}_{1}\left(  \alpha|Z\right)  -\widehat{\Phi}_{1,k+1}\left[
\gamma_{1}\left(  \alpha\right)  Z_{1},\gamma_{2}\left[  \widehat{G}%
_{2}\widehat{B}_{1}\left(  \alpha|Z\right)  \right]  Z_{2}\right]  \right)
^{2}d\alpha\right\}  dZ.\label{Gamit}%
\end{align}
Alternatively to (\ref{Gamit}), $\widehat{\gamma}_{11,k+1}\left(
\cdot\right)  $ and $\widehat{\gamma}_{12,k+1}\left(  \cdot\right)  $ can be
obtained by solving the differential system in Section \ref{2bc} using
$\widehat{\Phi}_{1,k+1}\left(  \cdot\right)  $ in place of $\Phi_{1}\left(
\cdot\right)  $. The stopping criterion must take into account that $\Phi
_{1}\left(  x_{1},x_{2}\right)  $ may depend only upon $x_{1}$ or $x_{2}$. For
instance, in the private value case $\Phi_{1}\left(  x_{1},x_{2}\right)
=\Phi_{1}\left(  x_{1}\right)  $, the valuation function is $\Phi_{1}\left[
\gamma_{11}\left(  \alpha\right)  Z_{1}\right]  $ and $\gamma_{12}\left(
\cdot\right)  $ is not identified. If this holds, $\widehat{\gamma}%
_{12,k}\left(  \cdot\right)  $ may not converge when $k$ grows. This can be
addressed by dropping out of the minimization the corresponding slope when the
sieve coefficients of $\widehat{\Phi}_{1,k}\left(  x_{1},x_{2}\right)  $ shows
that this function may not depend upon $x_{1}$ or $x_{2}$.

\setcounter{equation}{0}

\section{Conclusion\label{conclusion}}

The present paper considers a nonparametric interdependent value model which
is shown to be identified from first-price auction best response bids. The
model is derived from Milgrom and Weber (1982) and assumes that the bidder
signal depends upon some observed bidder characteristics, which variations are
key to obtain identification. Compared to other approaches of the literature,
this specification does not rely on functional restrictions difficult to
maintain or to test and delivers valuation functions that can be computed for
all possible signal values. The latter allows to implement various
counterfactuals, such as expected revenue computation for alternative auction
scenarii. Most of the conditions ensuring identification are testable. The
considered interdependent value model is overidentified, so that specification
testing is possible.

While the proposed approach assumes that the bidder private signal and information shifter are combined using a linear index structure, we believe that the identification procedure is general enough to tackle various other functional forms. 
The linear index may also be viewed as a nonparametric approximation of a function combining private signals and information shifter. Unobserved bidder heterogeneity would also deserve further research, investigating for instance implementations of nonparametric deconvolution techniques as in Li et al. (2000) or Krasnokutskaya (2011), the approach of Compiani et al. (2018) and Haile and Kitamura (2018), or parametric specifications.

\pagebreak

\section*{Appendix: Proofs of the results}

\renewcommand{\thesection}{A.\arabic{section}}
\renewcommand{\thetheorem}{A.\arabic{theorem}}
\renewcommand{\theequation}{A.\arabic{equation}}
\setcounter{section}{0}
\setcounter{equation}{0} \setcounter{theorem}{0} \setcounter{footnote}{0}

\section{Proof of Lemma \ref{Bidder2}}

As $B_{j}=s_{j}(A_{j};Z)$ where $s_{j}\left(  \cdot;Z\right)  $ is strictly
increasing under Assumption S-(ii) and because $A_{j}$ is a $\mathcal{U}%
_{\left[  0,1\right]  }$ random variable, it holds for all $b$ in $\left[
s_{j}(0;Z),s_{j}(1;Z)\right]  $%
\begin{align*}
G_{j}\left(  b|Z\right)   &  =\mathbb{P}\left(  B_{j}\leq b|Z\right)
=\mathbb{P}\left[  s_{j}(A_{j};Z)\leq b|Z\right]  =\mathbb{P}\left[  A_{j}\leq
s_{j}^{-1}(b;Z)|Z\right] \\
&  =s_{j}^{-1}(b;Z).
\end{align*}
Hence $G_{j}\left(  B_{j}|Z\right)  =s_{j}^{-1}\left[  s_{j}(A_{j}%
;Z);Z\right]  =A_{j}$ and $B_{j}\left(  \cdot|Z\right)  =G_{j}^{-1}\left(
\cdot|Z\right)  =s_{j}\left(  \cdot;Z\right)  $, which establish (i) and (ii).
(iii) follows from%
\[
\omega\left(  a|\alpha,Z\right)  =\mathbb{P}\left[  \left.  s_{i}\left(
a;Z\right)  >\max_{1\leq j\neq i\leq n}B_{j}\right\vert A_{i}=\alpha,Z\right]
\]
and $s_{i}\left(  \cdot;Z\right)  =B_{i}\left(  \cdot|Z\right)  $%
.$\hfill\square$

\section{Proof of Theorems \ref{MSMidentn2} and \ref{MSMidentn3}}

In this proof, we assume $i=1$ without loss of generality and remove the
corresponding index for the sake of brevity. In what follows $\left|\cdot\right|$ stands for the Euclidean norm of a vector or the absolute value of a real number. $C$ denotes a constant that may vary from line to line.

\subsection{The two bidder case: proof of Theorem \ref{MSMidentn2}}
We detail here the proof of Step 3 in Section \ref{2bc}. Recall it was shown in Section \ref{2bc} that $\Phi(\cdot)$ is identified over $\mathbb{R}_{+}^2$  from
\[
	U\left(  \alpha|Z\right)  =\Phi\left[  Z_{1}^{\prime}\gamma_{1}\left(
	\alpha\right)  ,Z_{2}^{\prime}\gamma_{2}\left(  G_{2}B_{1}\left(
	\alpha|Z\right)  \right)  \right]  
\]
which is also identified, as $\gamma_1^{} (1)$ and $\gamma_2 (1)$. We now show that $\gamma_1 (\cdot)$ and $\gamma_2 (\cdot)$ are identified.
Differentiating $U(\alpha|Z)$ with respect to $\alpha$ gives
\begin{align*}
	&  \frac{\partial\Phi}{\partial x_{1}}\left[  Z_{1}^{\prime}\gamma_{1}\left(
	\alpha\right)  ,Z_{2}^{\prime}\gamma_{2}\left[  G_{2}B_{1}\left(
	\alpha|Z\right)  \right]  \right]  \frac{d\left\{  Z_{1}^{\prime}\gamma
		_{1}\left(  \alpha\right)  \right\}  }{d\alpha}\\
	&  \quad+\frac{\partial\Phi}{\partial x_{2}}\left[  Z_{1}^{\prime}\gamma
	_{1}\left(  \alpha\right)  ,Z_{2}\gamma_{2}\left[  G_{2}B_{1}\left(
	\alpha|Z\right)  \right]  \right]  \frac{\partial\left\{  Z_{2}^{\prime}%
		\gamma_{2}\left[  G_{2}B_{1}\left(  \alpha|Z\right)  \right]  \right\}
	}{\partial\alpha}=\frac{\partial U\left(  \alpha|Z\right)  }{\partial\alpha}.
\end{align*}
Differentiating with respect to the entry $Z_{2d}\ $gives, by (\ref{Dz2da}),%
\begin{align*}
	&  \frac{\partial\Phi}{\partial x_{2}}\left[  Z_{1}^{\prime}\gamma_{1}\left(
	\alpha\right)  ,Z_{2}^{\prime}\gamma_{2}\left[  G_{2}B_{1}\left(
	\alpha|Z\right)  \right]  Z_{2}\right]  \times\gamma_{2d}\left[  G_{2}%
	B_{1}\left(  \alpha|Z\right)  \right] \\
	&  \quad+\frac{\partial\Phi}{\partial x_{2}}\left[  Z_{1}^{\prime}\gamma
	_{1}\left(  \alpha\right)  ,Z_{2}^{\prime}\gamma_{2}\left[  G_{2}B_{1}\left(
	\alpha|Z\right)  \right]  \right]  \frac{\frac{\partial G_{2}B_{1}\left(
			\alpha|Z\right)  }{\partial Z_{2d}}}{g_{2}b_{1}\left(  \alpha|Z\right)  }%
	\frac{\partial\left\{  Z_{2}^{\prime}\gamma_{2}\left[  G_{2}B_{1}\left(
		\alpha|Z\right)  \right]  \right\}  }{\partial\alpha}=\frac{\partial U\left(
		\alpha|Z\right)  }{\partial Z_{2d}}%
\end{align*}
which implies, for the column Gradient vector
$
\partial_{Z_2} U\left(  \alpha|Z\right) 
=
\left[
\frac{\partial U\left(
\alpha|Z\right)  }{\partial Z_{2d}}
\right]^{\prime}_{d=1,\ldots,D}
$,
\begin{align*}
	&  \frac{\partial\Phi}{\partial x_{2}}\left[  Z_{1}^{\prime}\gamma_{1}\left(
	\alpha\right)  ,Z_{2}^{\prime}\gamma_{2}\left[  G_{2}B_{1}\left(
	\alpha|Z\right)  \right]  Z_{2}\right]  \times Z_{2}^{\prime}\gamma_{2}\left[
	G_{2}B_{1}\left(  \alpha|Z\right)  \right] \\
	&  
	\quad
	+
	\frac{\partial\Phi}{\partial x_{2}}
	\left[  Z_{1}^{\prime}\gamma_{1} \left(  \alpha\right)  
	,
	Z_{2}^{\prime}\gamma_{2}\left[  G_{2}B_{1}\left(\alpha|Z\right)  \right]  
	\right]  
	\frac{Z_{2}^{\prime}\partial_ {Z_{2}}G_{2}B_{1}\left(  \alpha|Z\right)  }{g_{2}b_{1}\left(\alpha|Z\right)  }
	\frac{\partial\left\{  Z_{2}^{\prime}\gamma_{2}\left[
		G_{2}B_{1}\left(  \alpha|Z\right)  \right]  \right\}  }{\partial\alpha}%
	=Z_{2}^{\prime}\partial_{Z_2} U\left(  \alpha|Z\right) %
\end{align*}
Define now 
\[\Gamma\left(  \alpha|Z\right)  =\left[  Z_{1}^{\prime}\gamma
_{1}\left(  \alpha\right)  ,Z_{2}^{\prime}\gamma_{2}\left[  G_{2}B_{1}\left(
\alpha|Z\right)  \right]  ^{\prime}\right]  ^{\prime}
\] 
and for
\[
\frac{\partial\mathbf{\Phi}}{\partial x_{j}}\left[  \Gamma\right]  \left(
\alpha|Z\right)  =\frac{\partial\Phi}{\partial x_{j}}\left[  Z_{1}^{\prime
}\gamma_{1}\left(  \alpha\right)  ,Z_{2}^{\prime}\gamma_{2}\left[  G_{2}%
B_{1}\left(  \alpha|Z\right)  \right]  \right]
\]
consider the $2\times2$ matrices%
\[
\mathbf{D}\left[  \Phi,\Gamma\right]  \left(  \alpha|Z\right)  =\left[
\begin{array}
[c]{cc}%
\frac{\partial\mathbf{\Phi}}{\partial x_{1}}\left[  \Gamma\right]  \left(
\alpha|Z\right)  & 0\\
0 & \frac{\partial\mathbf{\Phi}}{\partial x_{2}}\left[  \Gamma\right]  \left(
\alpha|Z\right)
\end{array}
\right]  ,
\]
\[
\mathbf{G}_{2}\left(  \alpha|Z\right) 
 =
\left[
\begin{array}
[c]{cc}%
1 & 1\\
0 & 
\frac{Z_{2}^{\prime}\partial_{Z_{2}} G_{2}B_{1}\left(  \alpha|Z\right)}{g_{2}b_{1}\left(\alpha|Z\right)}  
\end{array}
\right] 
=
\left[
\begin{array}
[c]{cc}%
1 & 0\\
0 & 
\frac{\alpha (1-\alpha)}{g_{2}b_{1}\left(\alpha|Z\right)}  
\end{array}
\right] 
\times
\left[
\begin{array}
[c]{cc}%
1 & 1\\
0 & 
\frac{Z_{2}^{\prime}\partial_{Z_{2}} G_{2}B_{1}\left(  \alpha|Z\right)}{\alpha (1-\alpha)}  
\end{array}
\right] 
\]
so that Assumptions I and S-(iv) with Lemma \ref{Bidder2}-(ii) ensure that
$\mathbf{G}_{2}\left(  \alpha|Z\right)$ has an inverse when $\alpha$ belongs to $(0,1)$ with
\[
\mathbf{G}_{2}^{-1} 
\left(  \alpha|Z\right)  
=
\left[
\begin{array}
[c]{cc}%
1 
& 
-
\frac{g_{2}b_{1}\left(\alpha|Z\right)}{Z_{2}^{\prime}\partial_{Z_{2}} G_{2}B_{1}\left(  \alpha|Z\right)}
\\
0 
& 
\frac{g_{2}b_{1}\left(\alpha|Z\right)}{Z_{2}^{\prime}\partial_{Z_{2}} G_{2}B_{1}\left(  \alpha|Z\right)}
\end{array}
\right] 
=
\left[
\begin{array}
[c]{cc}%
1 & 
-\frac{\alpha (1-\alpha)}{Z_{2}^{\prime}\partial_{Z_{2}} G_{2}B_{1}\left(  \alpha|Z\right)} \\
0 & 
\frac{\alpha (1-\alpha)}{Z_{2}^{\prime}\partial_{Z_{2}} G_{2}B_{1}\left(  \alpha|Z\right)}  
\end{array}
\right]
\times
\left[
\begin{array}
[c]{cc}%
1 & 0\\
0 & 
\frac{g_{2}b_{1}\left(\alpha|Z\right)}{\alpha (1-\alpha)}  
\end{array}
\right] 
.
\]
Observe that $\mathbf{D}\left[  \Phi,\Gamma\right]  \left(  \alpha|Z\right)$ has an inverse for all $(\alpha,Z)$ by Assumption P, while Assumption I gives that $\mathbf{G}_{2}\left(  \alpha|Z\right)$ diverges with the order $1/(1-\alpha)$ when $\alpha$ goes to $1$.
Define also the vector%
\[
\mathbf{\Psi}\left[  \Phi,\Gamma\right]  \left(  \alpha|Z\right)  =\left[
\begin{array}
[c]{c}%
\partial_{Z_2} U\left(  \alpha|Z\right)  \\
g_{2}b_{1}\left(  \alpha|Z\right)  
\left\{  
	Z_{2}^{\prime}\partial_{Z_2} U\left(  \alpha|Z\right) 
	-\frac{\partial\mathbf{\Phi}}{\partial x_{2}}
	\left[  \Gamma\right]  \left(  \alpha|Z\right)  
	Z_{2}^{\prime}\gamma_{2}\left[  G_{2}B_{1}\left(  \alpha|Z\right)  \right]
\right\}
\end{array}
\right]  .
\]
Then the differential system above writes $\mathbf{G}_{2}\left(
\alpha|Z\right)  \mathbf{D}\left[  \Phi,\Gamma\right]  \left(  \alpha
|Z\right)  \Gamma^{(1)} \left(  \alpha|Z\right)  =\mathbf{\Psi
}\left[  \Phi,\Gamma\right]  \left(  \alpha|Z\right)  $, so that
$\Gamma\left(  \cdot|Z\right)  $ must solve
\begin{equation}
\Gamma^{(1)} \left(  \alpha|Z\right)  
=
\left\{  
\mathbf{G}_{2}\left(  \alpha|Z\right)  
\mathbf{D}\left[  \Phi,\Gamma\right]  
\left(
\alpha|Z\right)  
\right\}^{-1}
\mathbf{\Psi}\left[  \Phi,\Gamma\right]
\left(  \alpha|Z\right)
\label{Diffsys}
\end{equation}
over $(0,1)$. As the LHS is continuous over $[0,1]$, so must be the RHS. Hence the differential system (\ref{Diffsys}) holds over $[0,1]$ 
with a known terminal value $\Gamma\left(
1|Z\right)  $ by Assumption S-(iii).

Suppose now that a continuously differentiable
$
\widetilde{\Gamma}\left(  \alpha|Z\right)  
=
\left[  
Z_{1}^{\prime}\widetilde{\gamma}_{1}\left(  \alpha\right)  
,
Z_{2}^{\prime} \widetilde{\gamma}_{2}\left[  G_{2}B_{1}\left(
\alpha|Z\right)  \right] ^{\prime}\right]  ^{\prime}
$
also solves the differential system (\ref{Diffsys}) with the terminal condition
$
\widetilde{\Gamma}\left(  1 |Z\right)
=
\Gamma \left(  1 |Z\right)
$.
Then (\ref{Diffsys}) gives
\begin{eqnarray}
\widetilde{\Gamma}^{(1)} (\alpha|Z)
-
\Gamma^{(1)} (\alpha|Z)
& = &
\left\{   
\mathbf{G}_{2} \left(  \alpha|Z\right)
\mathbf{D}\left[  \Phi,\widetilde{\Gamma}\right]  
\left(
\alpha|Z\right)  
\right\}^{-1}
\mathbf{\Psi}\left[  \Phi,\widetilde{\Gamma}\right]
\left(  \alpha|Z\right)
\nonumber \\
&& 
-
\left\{  
\mathbf{G}_{2}\left(  \alpha|Z\right)  
\mathbf{D}\left[  \Phi,\Gamma\right]  
\left(
\alpha|Z\right)  
\right\}^{-1}
\mathbf{\Psi}\left[  \Phi,\Gamma\right]
\left(  \alpha|Z\right)
.
\label{Dgam1}
\end{eqnarray}
Let $\left\vert \cdot\right\vert $ be the Euclidean norm
and set, for a fixed $Z$,
$
\Delta (\alpha) 
=
\widetilde{\Gamma} (\alpha|Z)
-
\Gamma (\alpha|Z) 
$. Note that $\Delta (\cdot)$ is continuously differentiable with $\Delta (1) =0$, so that there exists a $\lambda>0$ such that for all $\alpha$
\[
\left\vert \Delta (\alpha)\right\vert
\leq
\lambda (1-\alpha).
\] 
Since the partial derivatives of $\Phi (\cdot)$ are Lipshitz by Assumption P, the expression of
$\mathbf{G}_{2}^{-1} 
\left(  \alpha|Z\right)$
and Assumption I imply by (\ref{Dgam1}), for all $\alpha$ in $[0,1]$,
\[
\left\vert 
\Delta^{(1)} (\alpha)
\right\vert
\leq
\frac{C}{1-\alpha}
\left\vert 
\Delta (\alpha)
\right\vert
.
\]
Hence for all $\epsilon>0$ small enough, (\ref{Dgam1}) gives
\begin{eqnarray}
\left\vert 
\Delta^{(1)} (\alpha)
\right\vert
& \leq &
\mathbb{I}
\left(
\alpha \leq 1- \epsilon
\right)
	\frac{C
	}
	{\epsilon}
	\left\vert 
	\Delta (\alpha)
	\right\vert
	+
	\mathbb{I}
	\left(
	1- \epsilon < \alpha \leq 1 
	\right)
	\lambda
\label{Dgam2}.
\end{eqnarray}

It follows from (\ref{Dgam2}) that
\[
\left\vert 
\Delta^{(1)} (\alpha)
\right\vert
\leq
\mathbb{I}
\left(
\alpha \leq 1- \epsilon
\right)
\lambda
\frac{C 
	}
{\epsilon}  (1-\alpha)
+
\mathbb{I}
\left(
1- \epsilon < \alpha \leq 1 
\right)
\lambda
\]
and then, since $\Delta (1)=0$,
\begin{eqnarray*}
\left|
\Delta (\alpha)
\right|
& = &
\left|
\int_{\alpha}^{1}
\Delta^{(1)} (t)
dt
\right|
\leq
\mathbb{I}
\left(
\alpha \leq 1- \epsilon
\right)
\lambda
\frac{C 
	}
{\epsilon}
\frac{(1-\alpha)^2}{2}
+
\mathbb{I}
\left(
1- \epsilon < \alpha \leq 1 
\right)
\lambda
(1-\alpha)
\\
& \leq &
\mathbb{I}
\left(
\alpha \leq 1- \epsilon
\right)
\lambda
\frac{C 
}
{\epsilon}
\frac{(1-\alpha)^2}{2}
+
\mathbb{I}
\left(
1- \epsilon < \alpha \leq 1 
\right)
\lambda
\epsilon
.
\end{eqnarray*}
Substituting in (\ref{Dgam2}) shows that
\begin{eqnarray*}
\left\vert 
\Delta^{(1)} (\alpha)
\right\vert
\leq
\mathbb{I}
\left(
\alpha \leq 1- \epsilon
\right)
\lambda
\left(
\frac{C 
}
{\epsilon}
\right)^2
\frac{(1-\alpha)^2}{2}
+
\mathbb{I}
\left(
1- \epsilon < \alpha \leq 1 
\right)
\lambda.
\end{eqnarray*}
An additional iteration shows that
\begin{eqnarray*}
\left|
\Delta (\alpha)
\right|
\leq 
\mathbb{I}
\left(
\alpha \leq 1- \epsilon
\right)
\lambda
\left(
\frac{C 
}
{\epsilon}
\right)^2
\frac{(1-\alpha)^3}{3!}
+
\mathbb{I}
\left(
1- \epsilon < \alpha \leq 1 
\right)
\lambda
\epsilon,
\\
\left|
\Delta^{(1)} (\alpha)
\right|
\leq 
\mathbb{I}
\left(
\alpha \leq 1- \epsilon
\right)
\lambda
\left(
\frac{C 
}
{\epsilon}
\right)^3
\frac{(1-\alpha)^3}{3!}
+
\mathbb{I}
\left(
1- \epsilon < \alpha \leq 1 
\right)
\lambda
.
\end{eqnarray*}
Iterating then gives, for any integer number $p\geq 2$, any $\epsilon>0$ small enough and all $\alpha$ of $[0,1]$,
\begin{eqnarray*}
\left|
\Delta (\alpha)
\right|
\leq 
\mathbb{I}
\left(
\alpha \leq 1- \epsilon
\right)
\lambda
\frac{\epsilon}{C}
\left(
\frac{C 
}
{\epsilon}
\right)^p
\frac{(1-\alpha)^p}{p!}
+
\mathbb{I}
\left(
1- \epsilon < \alpha \leq 1 
\right)
\lambda
\epsilon
\end{eqnarray*}
As
$
\lim_{p\uparrow\infty}  \left(
\frac{C 
}
{\epsilon}
\right)^p
\frac{(1-\alpha)^p}{p!}
=
0
$,
it follows that for all $\alpha$ in $[0,1]$,
$\left|
\Delta (\alpha)
\right|
\leq 
\lambda \epsilon$
for all $\epsilon>0$, which implies $\Delta (\cdot) = 0$ over $[0,1]$. 
Then  Assumption S and Lemma \ref{Bidder2}-(ii) imply that
$\widetilde{\gamma}_1 (\cdot) = \gamma_1 (\cdot)$ and $\widetilde{\gamma}_2 (\cdot) = \gamma_2 (\cdot)$. Hence the slope $[\gamma_1 (\cdot),\gamma_2 (\cdot)]$ is identified. This ends the proof of the Theorem. $\hfill\square$

\subsection{More than two bidders\label{Mttb}: proof of Theorem \ref{MSMidentn3}}

For the sake of notation, assume $n=3$, the case of a larger number of bidders
being similar, and set $i=1$.

\paragraph{Identification of $\Phi(\cdot)$ and $\gamma_j(0)$, $j$ in $\mathcal{I}$.}
Let
\[
W\left(  \alpha|Z\right)  =\left.  \frac{\partial\overline{V}_{1}\left(
	a|\alpha,Z\right)  }{\partial a}\right\vert _{a=\alpha}%
\]
be the function in (\ref{U1I3}), which is%
\begin{align*}
&  W\left(  \alpha|Z\right)  =g_{3}b_{1}\left(  \alpha|Z\right)  \int%
_{0}^{G_{2}B_{1}\left(  \alpha|Z\right)  }\Phi\left[  Z_{1}^{\prime}\gamma
_{1}\left(  \alpha\right)  ,Z_{2}^{\prime}\gamma_{2}\left(  t_{2}\right)
,Z_{3}^{\prime}\gamma_{3}\left[  G_{3}B_{1}\left(  \alpha|Z\right)  \right]
\right]  c\left(  t_{2}|\alpha,Z\right)  dt_{2}\\
&  \quad+g_{2}b_{1}\left(  \alpha|Z\right)  \int_{0}^{G_{3}B_{1}\left(
	\alpha|Z\right)  }\Phi\left[  Z_{1}^{\prime}\gamma_{1}\left(  \alpha\right)
,Z_{2}^{\prime}\gamma_{2}\left[  G_{2}B_{1}\left(  \alpha|Z\right)  \right]
,Z_{3}^{\prime}\gamma_{3}\left(  t_{3}\right)  \right]  c\left(  t_{3}%
|\alpha,Z\right)  dt_{3}%
\end{align*}
setting $c\left(  t_{i}|\alpha,Z\right)  =c\left(  t_{i}|A_{1}=\alpha
,A_{j}=G_{j}B_{1}\left(  \alpha|Z\right)  ,Z\right)  $ where $\left(
i,j\right)  $ is $\left(  2,3\right)  $ or $\left(  3,2\right)  $. With this
notation%
\begin{align*}
\left.  \frac{\partial\omega\left(  a|\alpha,Z\right)  }{\partial
	a}\right\vert _{a=\alpha}  &  =g_{3}b_{1}\left(  \alpha|Z\right)  \int%
_{0}^{G_{2}B_{1}\left(  \alpha|Z\right)  }c\left(  t_{2}|\alpha,Z\right)
dt_{2}\\
&  +g_{2}b_{1}\left(  \alpha|Z\right)  \int_{0}^{G_{3}B_{1}\left(
	\alpha|Z\right)  }c\left(  t_{3}|\alpha,Z\right)  dt_{3},\\
U_1\left(  \alpha|Z\right)   &  =\frac{W\left(  \alpha|Z\right)  }{\left.
	\frac{\partial\omega\left(  a|\alpha,Z\right)  }{\partial a}\right\vert
	_{a=\alpha}}.
\end{align*}
This implies by the initial bid condition in Assumption S-(ii) and by S-(iv),%
\[
U_1\left(  0|Z\right)  =\lim_{\alpha\downarrow0}\frac{W\left(  \alpha|Z\right)
}{\left.  \frac{\partial\omega\left(  a|\alpha,Z\right)  }{\partial
		a}\right\vert _{a=\alpha}}=\Phi\left[  Z_{1}^{\prime}\gamma_{1}\left(
0\right)  ,Z_{2}^{\prime}\gamma_{2}\left(  0\right)  ,Z_{3}^{\prime}\gamma
_{3}\left(  0\right)  \right]  .
\]
This identifies the set $\mathcal{I}$ of active signals and the corresponding
$\gamma_{j}\left(  0\right)  $ through the partial derivatives $\partial_{Z_{j}} U_1 (0|Z)$, Assumption G-(ii),
Assumption P and using the normalization condition 
$\frac{\partial}{\partial x_j} \Phi(0,0,0)=1$. Assumptions Z and G-(ii) ensures it is possible to find $Z_{1}$,
$Z_{2\,}$ and $Z_{3}$ such that $\left(  Z_{1}^{\prime}\gamma_{1}\left(
0\right)  ,Z_{2}^{\prime}\gamma_{2}\left(  0\right)  ,Z_{3}^{\prime}\gamma
_{3}\left(  0\right)  \right)  =\left(  x_{1},x_{2},x_{3}\right)  $ for any
$\left(  x_{1},x_{2},x_{3}\right)  $ in $\mathbb{R}_{+*}^3$. This shows that $U_1\left(  0|Z\right)  $
identifies $\Phi\left(  \cdot\right)  $ over $\mathbb{R}_{+}^{3}$ by continuity.

\paragraph{Identification of $\gamma_j(\cdot)$, $j$ in $\mathcal{I}$.}
Suppose all the signals are active, $\mathcal{I=}\left\{  1,2,3\right\}  $,
the other cases being similar. As for the two bidder case, the proof proceeds
by finding an integro-differential system which unique solution is $\left(
\gamma_{1}\left(  \cdot\right)  ,\gamma_{2}\left(  \cdot\right)  ,\gamma
_{3}\left(  \cdot\right)  \right)  $. Set%
\[
\Gamma\left(  \alpha|Z\right)  =\left[  \gamma_{1}\left(  \alpha\right)
,\gamma_{2}\left[  G_{2}B_{1}\left(  \alpha|Z\right)  \right]  ,\gamma
_{3}\left[  G_{3}B_{1}\left(  \alpha|Z\right)  \right]  \right]  ^{\prime}.
\]
Differentiating $W\left(  \alpha|Z\right)  $ with respect to $\alpha$ gives%
\begin{align*}
&  \mathbf{D}_{1}\left[  \Phi,\Gamma\right]  \left(  \alpha|Z\right)
\frac{d\left\{  Z_{1}^{\prime}\gamma_{1}\left(  \alpha\right)  \right\}
}{d\alpha}+g_{2}b_{1}\left(  \alpha|Z\right)  \mathbf{D}_{2}\left[
\Phi,\Gamma\right]  \left(  \alpha|Z\right)  \frac{\partial\left\{
Z_{2}^{\prime}\gamma_{2}\left[  G_{2}B_{1}\left(  \alpha|Z\right)  \right]
\right\}  }{\partial\alpha}\\
&  \quad+g_{3}b_{1}\left(  \alpha|Z\right)  \mathbf{D}_{3}\left[  \Phi
,\Gamma\right]  \left(  \alpha|Z\right)  \frac{\partial\left\{  Z_{3}^{\prime
}\gamma_{3}\left[  G_{3}B_{1}\left(  \alpha|Z\right)  \right]  \right\}
}{\partial\alpha}=\mathbf{\Psi}_{1}\left[  \Phi,\Gamma\right]  \left(
\alpha|Z\right)
\end{align*}
where%
\begin{align*}
\mathbf{D}_{1}\left[  \Phi,\Gamma\right]  \left(  \alpha|Z\right)   &
=g_{3}b_{1}\left(  \alpha|Z\right)  \int_{0}^{G_{2}B_{1}\left(  \alpha
|Z\right)  }\frac{\partial\Phi}{\partial x_{1}}\left[  Z_{1}^{\prime}%
\gamma_{1}\left(  \alpha\right)  ,Z_{2}^{\prime}\gamma_{2}\left(
t_{2}\right)  ,Z_{3}^{\prime}\gamma_{3}\left[  G_{3}B_{1}\left(
\alpha|Z\right)  \right]  \right]  c\left(  t_{2}|\alpha,Z\right)  dt_{2}\\
&  +g_{2}b_{1}\left(  \alpha|Z\right)  \int_{0}^{G_{3}B_{1}\left(
\alpha|Z\right)  }\frac{\partial\Phi}{\partial x_{1}}\left[  Z_{1}^{\prime
}\gamma_{1}\left(  \alpha\right)  ,Z_{2}^{\prime}\gamma_{2}\left[  G_{2}%
B_{1}\left(  \alpha|Z\right)  \right]  ,Z_{3}^{\prime}\gamma_{3}\left(
t_{3}\right)  \right]  c\left(  t_{3}|\alpha,Z\right)  dt_{3},\\
\mathbf{D}_{2}\left[  \Phi,\Gamma\right]  \left(  \alpha|Z\right)   &
=\int_{0}^{G_{3}B_{1}\left(  \alpha|Z\right)  }\frac{\partial\Phi}{\partial
x_{2}}\left[  Z_{1}^{\prime}\gamma_{1}\left(  \alpha\right)  ,Z_{2}^{\prime
}\gamma_{2}\left[  G_{2}B_{1}\left(  \alpha|Z\right)  \right]  ,Z_{3}^{\prime
}\gamma_{3}\left(  t_{3}\right)  \right]  c\left(  t_{3}|\alpha,Z\right)
dt_{3},\\
\mathbf{D}_{3}\left[  \Phi,\Gamma\right]  \left(  \alpha|Z\right)   &
=\int_{0}^{G_{2}B_{1}\left(  \alpha|Z\right)  }\frac{\partial\Phi}{\partial
x_{3}}\left[  Z_{1}^{\prime}\gamma_{1}\left(  \alpha\right)  ,Z_{2}^{\prime
}\gamma_{2}\left(  t_{2}\right)  ,Z_{3}^{\prime}\gamma_{3}\left[  G_{3}%
B_{1}\left(  \alpha|Z\right)  \right]  \right]  c\left(  t_{2}|\alpha
,Z\right)  ,
\end{align*}
and where $\mathbf{\Psi}_{1}\left[  \Phi,\Gamma\right]  \left(  \alpha
|Z\right)  $ is equal to%
\begin{align*}
&  \frac{\partial W\left(  \alpha|Z\right)  }{\partial\alpha}-\frac{\partial
g_{2}b_{1}\left(  \alpha|Z\right)  }{\partial\alpha}\int_{0}^{G_{3}%
B_{1}\left(  \alpha|Z\right)  }\Phi\left[  Z_{1}^{\prime}\gamma_{1}\left(
\alpha\right)  ,Z_{2}^{\prime}\gamma_{2}\left[  G_{2}B_{1}\left(
\alpha|Z\right)  \right]  ,Z_{3}^{\prime}\gamma_{3}\left(  t_{3}\right)
\right]  c\left(  t_{3}|\alpha,Z\right)  dt_{3}\\
&  \quad-\frac{\partial g_{3}b_{1}\left(  \alpha|Z\right)  }{\partial\alpha
}\int_{0}^{G_{2}B_{1}\left(  \alpha|Z\right)  }\Phi\left[  Z_{1}^{\prime
}\gamma_{1}\left(  \alpha\right)  ,Z_{2}^{\prime}\gamma_{2}\left(
t_{2}\right)  ,Z_{3}^{\prime}\gamma_{3}\left[  G_{3}B_{1}\left(
\alpha|Z\right)  \right]  \right]  c\left(  t_{2}|\alpha,Z\right)  dt_{2}\\
&  \quad-2g_{2}b_{1}\left(  \alpha|Z\right)  g_{3}b_{1}\left(  \alpha
|Z\right)  c\left[  G_{2}B_{1}\left(  \alpha|Z\right)  ,G_{3}B_{1}\left(
\alpha|Z\right)  |\alpha,Z\right] \\
&  \quad\quad\quad\quad\quad\quad\times\Phi\left[  Z_{1}^{\prime}\gamma
_{1}\left(  \alpha\right)  ,Z_{2}^{\prime}\gamma_{2}\left[  G_{2}B_{1}\left(
\alpha|Z\right)  \right]  ,Z_{3}^{\prime}\gamma_{3}\left[  G_{3}B_{1}\left(
\alpha|Z\right)  \right]  \right] \\
&  \quad-g_{2}b_{1}\left(  \alpha|Z\right)  \int_{0}^{G_{3}B_{1}\left(
\alpha|Z\right)  }\Phi\left[  Z_{1}^{\prime}\gamma_{1}\left(  \alpha\right)
,Z_{2}^{\prime}\gamma_{2}\left[  G_{2}B_{1}\left(  \alpha|Z\right)  \right]
,Z_{3}^{\prime}\gamma_{3}\left(  t_{3}\right)  \right]  \frac{\partial
c\left(  t_{3}|\alpha,Z\right)  }{\partial\alpha}dt_{3}\\
&  \quad-g_{3}b_{1}\left(  \alpha|Z\right)  \int_{0}^{G_{2}B_{1}\left(
\alpha|Z\right)  }\Phi\left[  Z_{1}^{\prime}\gamma_{1}\left(  \alpha\right)
,Z_{2}^{\prime}\gamma_{2}\left(  t_{2}\right)  ,Z_{3}^{\prime}\gamma
_{3}\left[  G_{3}B_{1}\left(  \alpha|Z\right)  \right]  \right]
\frac{\partial c\left(  t_{2}|\alpha,Z\right)  }{\partial\alpha}dt_{2}.
\end{align*}
Differentiating $W\left(  \alpha|Z\right)  $ with respect to $Z_{2d}$ gives,
by (\ref{Dz2da})%
\begin{align*}
&  
\mathbf{D}_{2}\left[  \Phi,\Gamma\right]  \left(  \alpha|Z\right)
\frac{1}{g_{2}b_{1}\left(\alpha|Z\right)}
\frac{\partial G_{2}B_{1}\left(  \alpha|Z\right)  }{\partial Z_{2d}
}
\frac{\partial\left\{  Z_{2}^{\prime}\gamma_{2}\left[  G_{2}B_{1}\left(
\alpha|Z\right)  \right]  \right\}  }{\partial\alpha}\\
&  \quad +
\mathbf{D}_{3}\left[  \Phi,\Gamma\right]  \left(  \alpha|Z\right)
\frac{1}{g_{3}b_{1}\left(
	\alpha|Z\right)}
\frac{\partial G_{3}B_{1}\left(  \alpha|Z\right)  }{\partial Z_{2d}%
}
\frac{\partial\left\{  Z_{3}^{\prime}\gamma_{3}\left[  G_{3}B_{1}\left(
\alpha|Z\right)  \right]  \right\}  }{\partial\alpha}=\mathbf{\psi}%
_{2d}\left[  \Phi,\Gamma\right]  \left(  \alpha|Z\right)
\end{align*}
where%
\begin{align*}
&  \mathbf{\psi}_{2d}\left[  \Phi,\Gamma\right]  \left(  \alpha|Z\right)
=\frac{\partial W\left(  \alpha|Z\right)  }{\partial Z_{2d}}\\
&  \quad-g_{2}b_{1}\left(  \alpha|Z\right)  \gamma_{2d}\left[  G_{2}%
B_{1}\left(  \alpha|Z\right)  \right]  \int_{0}^{G_{3}B_{1}\left(
\alpha|Z\right)  }\Phi_{x_{2}}\left[  Z_{1}^{\prime}\gamma_{1}\left(
\alpha\right)  ,Z_{2}^{\prime}\gamma_{2}\left[  G_{2}B_{1}\left(
\alpha|Z\right)  \right]  ,Z_{3}^{\prime}\gamma_{3}\left(  t_{3}\right)
\right]  c\left(  t_{3}|\alpha,Z\right)  dt_{3}\\
&  \quad-\frac{\partial g_{2}b_{1}\left(  \alpha|Z\right)  }{\partial Z_{2d}%
}\int_{0}^{G_{3}B_{1}\left(  \alpha|Z\right)  }\Phi\left[  Z_{1}^{\prime
}\gamma_{1}\left(  \alpha\right)  ,Z_{2}^{\prime}\gamma_{2}\left[  G_{2}%
B_{1}\left(  \alpha|Z\right)  \right]  ,Z_{3}^{\prime}\gamma_{3}\left(
t_{3}\right)  \right]  c\left(  t_{3}|\alpha,Z\right)  dt_{3}\\
&  \quad-\frac{\partial g_{3}b_{1}\left(  \alpha|Z\right)  }{\partial Z_{2d}%
}\int_{0}^{G_{2}B_{1}\left(  \alpha|Z\right)  }\Phi\left[  Z_{1}^{\prime
}\gamma_{1}\left(  \alpha\right)  ,Z_{2}^{\prime}\gamma_{2}\left(
t_{2}\right)  ,Z_{3}^{\prime}\gamma_{3}\left[  G_{3}B_{1}\left(
\alpha|Z\right)  \right]  \right]  c\left(  t_{2}|\alpha,Z\right)  dt_{2}\\
&  \quad-\left(  g_{2}b_{1}\left(  \alpha|Z\right)  \frac{\partial G_{3}%
B_{1}\left(  \alpha|Z\right)  }{\partial Z_{2d}}+g_{3}b_{1}\left(
\alpha|Z\right)  \frac{\partial G_{2}B_{1}\left(  \alpha|Z\right)  }{\partial
Z_{2d}}\right) \\
&  \quad\quad\quad\times\Phi\left[  Z_{1}^{\prime}\gamma_{1}\left(
\alpha\right)  ,Z_{2}^{\prime}\gamma_{2}\left[  G_{2}B_{1}\left(
\alpha|Z\right)  \right]  ,Z_{3}^{\prime}\gamma_{3}\left[  G_{3}B_{1}\left(
\alpha|Z\right)  \right]  \right]  c\left[  G_{2}B_{1}\left(  \alpha|Z\right)
,G_{3}B_{1}\left(  \alpha|Z\right)  |\alpha\right] \\
&  \quad-g_{2}b_{1}\left(  \alpha|Z\right)  \int_{0}^{G_{3}B_{1}\left(
\alpha|Z\right)  }\Phi\left[  Z_{1}^{\prime}\gamma_{1}\left(  \alpha\right)
,Z_{2}^{\prime}\gamma_{2}\left[  G_{2}B_{1}\left(  \alpha|Z\right)  \right]
,Z_{3}^{\prime}\gamma_{3}\left(  t_{3}\right)  \right]  \frac{\partial
c\left(  t_{3}|\alpha,Z\right)  }{\partial Z_{2d}}dt_{3}\\
&  \quad-g_{3}b_{1}\left(  \alpha|Z\right)  \int_{0}^{G_{2}B_{1}\left(
\alpha|Z\right)  }\Phi\left[  Z_{1}^{\prime}\gamma_{1}\left(  \alpha\right)
,Z_{2}^{\prime}\gamma_{2}\left(  t_{2}\right)  ,Z_{3}^{\prime}\gamma
_{3}\left[  G_{3}B_{1}\left(  \alpha|Z\right)  \right]  \right]
\frac{\partial c\left(  t_{2}|\alpha,Z\right)  }{\partial Z_{2d}}dt_{2}.
\end{align*}
Multiplying by $Z_{2d}$ and summing then gives, for $\mathbf{\Psi}_{2}\left[
\Phi,\Gamma\right]  \left(  \alpha|Z\right)  =\sum_{d=1}^{D}Z_{2d}%
\mathbf{\psi}_{2d}\left[  \Phi,\Gamma\right]  \left(  \alpha|Z\right)  $,%
\begin{align*}
&  
\mathbf{D}_{2}\left[  \Phi,\Gamma\right]  \left(  \alpha|Z\right)
\frac{Z_{2}^{\prime} \partial_{Z_2} G_{2}B_{1}\left(  \alpha|Z\right)  }{g_{2}b_{1}\left(
	\alpha|Z\right)}
\frac{\partial\left\{  Z_{2}^{\prime}\gamma_{2}\left[  G_{2}B_{1}\left(
\alpha|Z\right)  \right]  \right\}  }{\partial\alpha}+\\
&  \quad 
\mathbf{D}_{3}\left[  \Phi,\Gamma\right]  \left(
\alpha|Z\right)  
\frac{Z_{2}^{\prime} \partial_{Z_2} G_{3}B_{1}\left(  \alpha|Z\right)
}{g_{3}b_{1}\left(  \alpha|Z\right)}
\frac{\partial\left\{  Z_{3}^{\prime}\gamma_{3}\left[
G_{3}B_{1}\left(  \alpha|Z\right)  \right]  \right\}  }{\partial\alpha
}=\mathbf{\Psi}_{2}\left[  \Phi,\Gamma\right]  \left(  \alpha|Z\right)
\end{align*}
A similar equation holds for $Z_{3}$. Let $\mathbf{D}\left[  \Phi
,\Gamma\right]  \left(  \alpha|Z\right)  $ be the $3\times3$ diagonal matrix
with entries $\mathbf{D}_{k}\left[  \Phi,\Gamma\right]  \left(  \alpha
|Z\right)  $ and $\mathbf{\Psi}\left[  \Phi,\Gamma\right]  \left(
\alpha|Z\right)  $ be the $3\times1$ vector with entries $\mathbf{\Psi}%
_{k}\left[  \Phi,\Gamma\right]  \left(  \alpha|Z\right)  $, $k=1,2,3$. Define%
\begin{eqnarray*}
\mathbf{G}\left(  \alpha|Z\right)  
& = &
\left[
\begin{array}
[c]{ccc}%
1 & g_{2}b_{1}\left(  \alpha|Z\right)  & g_{3}b_{1}\left(  \alpha|Z\right) \\
0 
& 
\frac{Z_{2}^{\prime} \partial_{Z_2} G_{2}B_{1}\left(  \alpha|Z\right)  }{
 g_{2}b_{1}\left(  \alpha|Z\right)} 
& 
\frac{Z_{2}^{\prime} \partial_{Z_2} G_{3}B_{1}\left(  \alpha|Z\right)
}{g_{3}b_{1}\left(  \alpha|Z\right)}
\\
0 
& 
\frac{Z_{3}^{\prime} \partial_{Z_3} G_{2}B_{1}\left(  \alpha|Z\right)  }{
g_{2}b_{1}\left(  \alpha|Z\right)} 
& \frac{Z_{3}^{\prime} \partial_{Z_3} G_{3}B_{1}\left(  \alpha|Z\right)
}{g_{3}b_{1}\left(  \alpha|Z\right)}%
\end{array}
\right]  
\\
& = &
\left[
\begin{array}
[c]{ccc}%
1 & g_{2}b_{1}\left(  \alpha|Z\right)  & g_{3}b_{1}\left(  \alpha|Z\right) \\
0 
& 
\frac{Z_{2}^{\prime} \partial_{Z_2} G_{2}B_{1}\left(  \alpha|Z\right)  }{
	\alpha (1-\alpha)} 
& 
\frac{Z_{2}^{\prime} \partial_{Z_2} G_{3}B_{1}\left(  \alpha|Z\right)
}{\alpha (1-\alpha)}
\\
0 
& 
\frac{Z_{3}^{\prime} \partial_{Z_3} G_{2}B_{1}\left(  \alpha|Z\right)  }{
	\alpha (1-\alpha)} 
& \frac{Z_{3}^{\prime} \partial_{Z_3} G_{3}B_{1}\left(  \alpha|Z\right)
}{\alpha (1-\alpha)}%
\end{array}
\right]  
\times
\left[
\begin{array}
[c]{ccc}%
1 & 0  & 0 \\
0 
& 
\frac{\alpha (1-\alpha)  }{
	g_{2}b_{1}\left(  \alpha|Z\right)} 
& 
0
\\
0 
& 
0
& \frac{\alpha (1-\alpha)
}{g_{3}b_{1}\left(  \alpha|Z\right)}%
\end{array}
\right] .
\end{eqnarray*}
As in the two bidder case, Assumptions I and S-(iv) with Lemma \ref{Bidder2}-(ii) ensure that
$\mathbf{G} \left(  \alpha|Z\right)$ has an inverse when $\alpha$ belongs to $(0,1)$ with $\mathbf{G}^{-1} \left(  \alpha|Z\right)$ of order $1/\alpha$ when $\alpha$ goes to $0$.
Observe as well that $\mathbf{D}\left[  \Phi,\Gamma\right]  \left(  \alpha|z\right)  $
has an inverse for all $\alpha$ under Assumption I. $\mathbf{G}\left(
\alpha|Z\right)  $ also has an inverse for $\alpha$ in $\left(  0,1\right)  $
but not if $\alpha=0$ or $\alpha=1$. Stacking the equations above together
shows that%
\[
\mathbf{G}\left(  \alpha|Z\right)  \mathbf{D}\left[  \Phi,\Gamma\right]
\left(  \alpha|Z\right)  \frac{d}{d\alpha}\Gamma\left(  \alpha|z\right)
=\mathbf{\Psi}\left[  \Phi,\Gamma\right]  \left(  \alpha|Z\right)
\]
for all $\alpha$ in $\left[  0,1\right]  $. This also gives for all $\alpha$
in $\left[  0,1\right]  $%
\begin{equation*}
\frac{d}{d\alpha}\Gamma\left(  \alpha|Z\right)  =\left\{  \mathbf{G}\left(
\alpha|Z\right)  \mathbf{D}\left[  \Phi,\Gamma\right]  \left(  \alpha
|Z\right)  \right\}  ^{-1}\mathbf{\Psi}\left[  \Phi,\Gamma\right]  \left(
\alpha|Z\right)  \label{DSg}%
\end{equation*}
passing at the limit in the RHS for $\alpha=0$ or $1$.

As above, identification of the slope functions holds provided two continuously differentiable solutions
of (\ref{DSg}), $\Gamma\left(  \cdot|Z\right)  $ and $\widetilde{\Gamma
}\left(  \cdot|Z\right)  $ with $\widetilde{\Gamma}\left(  0|Z\right)
=\Gamma\left(  0|Z\right)  $, must be equal. 
Now it holds
\begin{align*}
\frac{d}{d\alpha}\Gamma\left(  \alpha|Z\right)  
-
\frac{d}{d\alpha}\widetilde{\Gamma}\left(  \alpha |Z\right)   &
=  \left\{  \mathbf{G}\left(  \alpha|Z\right)
\mathbf{D}\left[  \Phi,\Gamma\right]  \left(  \alpha|Z\right)  \right\}
^{-1}\mathbf{\Psi}\left[  \Phi,\Gamma\right]  \left(  \alpha|Z\right)
\\
&   -  \left\{  \mathbf{G}\left(  \alpha|Z\right)
\mathbf{D}\left[  \Phi,\widetilde{\Gamma}\right]  \left(  \alpha|Z\right)
\right\}  ^{-1}\mathbf{\Psi}\left[  \Phi,\widetilde{\Gamma}\right]  \left(
\alpha|Z\right)   
\end{align*}
with 
$ \left| \mathbf{D}\left[  \Phi,\Gamma\right]  \left(  \alpha|Z\right) 
- 
\mathbf{D}\left[  \Phi,\widetilde{\Gamma}\right]  \left(  \alpha|Z\right)
\right| \leq C \alpha$ 
and 
$ \left| \mathbf{\Psi} \left[  \Phi,\Gamma\right]  \left(  \alpha|Z\right) 
- 
\mathbf{\Psi}\left[  \Phi,\widetilde{\Gamma}\right]  \left(  \alpha|Z\right)
\right| \leq C \alpha$ for all $\alpha$. Set, for a fixed $Z$,
$
\Delta (\alpha) 
=
\widetilde{\Gamma} (\alpha|Z)
-
\Gamma (\alpha|Z) 
$. Note that $\Delta (\cdot)$ is continuously differentiable with $\Delta (0) =0$,
$\Delta (\alpha)=\int_0^{\alpha} \Delta^{(1)} (t) dt$ and that there exists a $\lambda>0$ such that for all $\alpha$
\[
\left\vert \Delta (\alpha)\right\vert
\leq
\lambda \alpha .
\] 
It also holds for all $\alpha$ in $[0,1]$
\[
\left\vert 
\Delta^{(1)} (\alpha)
\right\vert
\leq
\frac{C}{\alpha}
\left\vert 
\Delta (\alpha)
\right\vert
.
\]
Hence for all $\epsilon>0$ small enough, (\ref{Dgam1}) gives
\begin{eqnarray*}
\left\vert 
\Delta^{(1)} (\alpha)
\right\vert
& \leq &
\mathbb{I}
\left(
\alpha \leq \epsilon
\right)
\lambda
+
\mathbb{I}
\left(
\epsilon < \alpha \leq 1 
\right)
\frac{C
}
{\epsilon}
\left\vert 
\Delta (\alpha)
\right\vert
.
\end{eqnarray*}
Arguing as in the two bidder case gives that $\Delta (\cdot)=0$, which establishes identification of the 
$\gamma_j(\cdot)$, $j \in \mathcal{I}$. This ends the proof of the Theorem. $\hfill\square$

\bigskip
\end{document}